# SELLING A STOCK AT THE ULTIMATE MAXIMUM

By Jacques du Toit and Goran Peskir

*The University of Manchester*

Assuming that the stock price $Z = (Z_t)_{0 \le t \le T}$ follows a geometric Brownian motion with drift $\mu \in \mathbb{R}$ and volatility $\sigma > 0$, and letting $M_t = \max_{0 \le s \le t} Z_s$ for $t \in [0, T]$, we consider the optimal prediction problems

$$V_1 = \inf_{0 \le \tau \le T} \mathsf{E}\left(\frac{M_T}{Z_\tau}\right) \quad \text{and} \quad V_2 = \sup_{0 \le \tau \le T} \mathsf{E}\left(\frac{Z_\tau}{M_T}\right),$$

where the infimum and supremum are taken over all stopping times $\tau$ of $Z$. We show that the following strategy is optimal in the first problem: if $\mu \le 0$ stop immediately; if $\mu \in (0, \sigma^2)$ stop as soon as $M_t/Z_t$ hits a specified function of time; and if $\mu \ge \sigma^2$ wait until the final time $T$. By contrast we show that the following strategy is optimal in the second problem: if $\mu \le \sigma^2/2$ stop immediately, and if $\mu > \sigma^2/2$ wait until the final time $T$. Both solutions support and reinforce the widely held financial view that "one should sell bad stocks and keep good ones." The method of proof makes use of parabolic free-boundary problems and local time–space calculus techniques. The resulting inequalities are unusual and interesting in their own right as they involve the future and as such have a predictive element.

**1. Introduction.** Imagine an investor who owns a stock which he wishes to sell before time $T > 0$ so as to maximize his profit. The investor has to decide when to sell the stock. Naturally, he would like to sell when the stock price is at its maximal value over the interval $[0, T]$, but such a strategy is impractical since this information is only known at time $T$. What the investor would like to do at any time $t \in [0, T]$ is to use all the accumulated information to infer how close the stock price is to the ultimate maximum, and based on this decide whether he should sell or not. In the present paper we consider the question of *predicting the maximum* when the stock follows a *geometric Brownian motion*. Following the initial publication [5], this question has arisen independently within circles of researchers and practitioners; however, all attempts at deriving a complete solution have been unsuccessful until now. For other optimal prediction problems studied to date we refer to [2, 3, 4, 9, 13, 14, 16] (see also [12], Chapter VIII).

The purpose of the present paper is to present the solution to this problem. Let $Z = (Z_t)_{0 \leq t \leq T}$ denote a stock price with drift $\mu \in \mathbb{R}$ and volatility $\sigma > 0$ such that

$$(1.1) \qquad dZ_t = \mu Z_t \, dt + \sigma Z_t \, dB_t,$$

where $B = (B_t)_{0 \leq t \leq T}$ is a standard Brownian motion. Setting

$$(1.2) \qquad M_t = \max_{0 \leq s \leq t} Z_s$$

for $t \in [0, T]$ we see that $M_T$ is the largest profit the investor could make from the sale. It is clear that the investor's selling strategy must be a stopping time taking values in $[0, T]$; however, for any such strategy $\tau$ there are several ways of evaluating its performance. One could deem $\tau$ to be a "good" strategy if the expected ratio $\mathsf{E}(M_T/Z_\tau)$ is small, or if the expected ratio $\mathsf{E}(Z_\tau/M_T)$ is big. It could also be "good" if the expected weighted difference $\mathsf{E}(M_T - Z_\tau)^p$ is small for some $p > 0$, or indeed if the expected difference $\mathsf{E}|\theta - \tau|$ is small where $\theta = \inf\{t \in [0, T] \mid Z_t = M_T\}$ denotes the time at which $Z$ attains its maximal value. Optimizing each of these performance measures over all stopping times in $[0, T]$ will typically yield different results, and it is up to the investor to decide which performance measure is most appropriate to him.

In the present paper we will judge performance based on the ratio of $M_T$ to $Z_\tau$. This formulation is very natural and has the effect of stripping away the monetary value of the stock and focusing only on the underlying randomness. The ratio is unitless (or dimensionless) meaning that expensive stocks and cheap stocks are treated in the same way. However, here as well, one can examine either the ratio $M_T/Z_\tau$ or the ratio $Z_\tau/M_T$ and there is no reason a priori to prefer either. This leads to the optimal prediction problems

$$(1.3) \qquad V_1 = \inf_{0 \leq \tau \leq T} \mathsf{E}\left(\frac{M_T}{Z_\tau}\right),$$

$$(1.4) \qquad V_2 = \sup_{0 \leq \tau \leq T} \mathsf{E}\left(\frac{Z_\tau}{M_T}\right),$$

where the infimum and supremum are taken over all stopping times $\tau$ of $Z$. While these two problem formulations have arisen independently within circles of researchers and practitioners following [5], to the best of our knowledge the first to record them in the present form was Shiryaev (see [13],

page 488). Note that $V_1$ aims at penalizing small values of $Z_\tau$ in relation to the size of $M_T$, while $V_2$ rewards large values of $Z_\tau$ and represents the highest percentage of $M_T$ one can attain with an adapted selling strategy. Since the two problems are so similar, one would expect them to have similar solutions. It is therefore quite surprising to find that the two solutions are *very different* for an important set of parameters.

For the infimum formulation (1.3) the optimal strategy is as follows (Theorem 1): if $\mu \leq 0$ stop immediately; if $\mu \in (0, \sigma^2)$ stop as soon as $M_t/Z_t$ hits a specified function of time; and if $\mu \geq \sigma^2$ wait until the final time $T$. By contrast (and quite unexpectedly) the supremum formulation (1.4) has the following solution (Theorem 2): if $\mu \leq \sigma^2/2$ stop immediately, and if $\mu > \sigma^2/2$ wait until the final time $T$. This solution extends and reinforces a recent result by Shiryaev, Xu and Zhou (presented at Sydney's QMF conference in 2007) that when $\mu \leq 0$ in (1.4) it is optimal to stop immediately and when $\mu \geq \sigma^2$ it is optimal to wait until the final time (see [15]). Apart from resolving the problem when $\mu \in (0, \sigma^2)$, and revealing a "bang–bang" strategy at $\mu = \sigma^2/2$ (see Remarks 1 and 2), our proof is purely probabilistic.

Both formulations therefore reinforce the widely held financial view that one should sell bad stocks and keep good ones; however, they disagree somewhat on which stocks are "good." It is also interesting that the infimum formulation (1.3) has a more sophisticated strategy: dividing the maximum $M_T$ by the stock price $Z_\tau$ exposes and magnifies the small perturbations produced by the Brownian motion, whereas dividing $Z_\tau$ by $M_T$ effectively dampens them out. Both strategies are also quite different from the optimal stopping time in [2] where a standard Brownian motion with drift was considered.

The solution to the optimal prediction problem (1.3) is derived in Section 3, and the solution to the optimal prediction problem (1.4) is derived in Section 4. It is interesting to note that although the optimal stopping time in the latter case is trivial, the proof nonetheless requires some effort [the case $\mu \in (0, \sigma^2/2)$ being the most demanding]. The resulting inequalities (Theorems 2 and 3) are unusual and interesting in their own right as they involve the future and as such have a predictive element.

**2. Formulation of the problem.** We begin our exposition by formally introducing the setting and the problem to be studied. Let $B = (B_t)_{0 \leq t \leq T}$ be a standard Brownian motion defined on the probability space $(\Omega, \mathcal{F}, \mathsf{P})$, and for any $\mu \in \mathbb{R}$ and $\sigma > 0$, let $Z = (Z_t)_{0 \leq t \leq T}$ be the unique strong solution to the stochastic differential equation

$$dZ_t = \mu Z_t \, dt + \sigma Z_t \, dB_t, \tag{2.1}$$

where the initial value $Z_0 > 0$ is taken to be independent from $B$. It is well known that $Z$ defines a geometric Brownian motion which is given by

$$Z_t = Z_0 \exp(\sigma B_t + (\mu - \sigma^2/2)t) \tag{2.2}$$

for $t \in [0,T]$. Defining the maximum $M = (M_t)_{0 \leq t \leq T}$ of the process $Z$ by $M_t = \max_{0 \leq s \leq t} Z_s$ we see from (2.2) that

$$(2.3) \qquad M_t = Z_0 \exp\left(\sigma \max_{0 \leq s \leq t}\left(B_s + \left(\frac{\mu - \sigma^2/2}{\sigma}\right)s\right)\right)$$

for $t \in [0,T]$. With this in mind, fix $\lambda \in \mathbb{R}$ and let $B^\lambda = (B_t^\lambda)_{0 \leq t \leq T}$ denote the Brownian motion with drift $\lambda$ given by $B_t^\lambda = B_t + \lambda t$ for $t \in [0,T]$. Defining the process $S^\lambda = (S_t^\lambda)_{0 \leq t \leq T}$ by $S_t^\lambda = \max_{0 \leq s \leq t} B_s^\lambda$, it follows that $M_t = Z_0 \exp(\sigma S_t^\lambda)$ for $t \in [0,T]$ where the drift $\lambda$ is given by $\lambda = (\mu - \sigma^2/2)/\sigma$.

Consider the optimal prediction problem

$$(2.4) \qquad V_1 = \inf_{0 \leq \tau \leq T} \mathsf{E}\left(\frac{M_T}{Z_\tau}\right) = \inf_{0 \leq \tau \leq T} \mathsf{E}(e^{\sigma(S_T^\lambda - B_\tau^\lambda)}),$$

where the infimum is taken over all stopping times $\tau$ of $Z$ (or $B$ equivalently) and $\lambda = (\mu - \sigma^2/2)/\sigma$. (Note that this expression is unitless since the initial value of the stock $Z_0$ does not appear.) The gain process $(S_T^\lambda - B_t^\lambda)_{0 \leq t \leq T}$ in the optimization problem above is not adapted to the natural filtration $(\mathcal{F}_t^B)_{0 \leq t \leq T}$ of $B$ as $S_T^\lambda$ is only $\mathcal{F}_T^B$ measurable. This means that (2.4) falls outside the scope of standard optimal stopping theory. However, using the same approach as in [9] it is possible to reduce (2.4) to an equivalent optimization problem to which the standard techniques of optimal stopping for Markov processes (see, e.g., [12]) can be applied. To do this, recall (cf. [1] and [8]) that the distribution function of $S_t^\lambda$ is given explicitly by

$$(2.5) \qquad \mathsf{P}(S_t^\lambda \leq x) = \Phi\left(\frac{x - \lambda t}{\sqrt{t}}\right) - e^{2\lambda x}\Phi\left(\frac{-x - \lambda t}{\sqrt{t}}\right)$$

for all $(t,x) \in (0,\infty) \times \mathbb{R}_+$, where $\Phi(x) = \int_{-\infty}^x \varphi(z)\,dz$ denotes the distribution function of a standard normal random variable, and $\varphi(x) = (1/\sqrt{2\pi})e^{-x^2/2}$ denotes its density function for $x \in \mathbb{R}$.

LEMMA 1. *The optimal prediction problem (2.4) is equivalent to the standard optimal stopping problem*

$$(2.6) \qquad V_1 = \inf_{0 \leq \tau \leq T} \mathsf{E}(G(\tau, S_\tau^\lambda - B_\tau^\lambda)),$$

*where the infimum is taken over all stopping times $\tau$ of $B$ and $\lambda = (\mu - \sigma^2/2)/\sigma$. When $\mu \neq 0$ the function $G$ is given by*

$$G(t,x) = \mathsf{E}(e^{\sigma(x \vee S_{T-t}^\lambda)}) = e^{\sigma x} + \sigma \int_x^\infty e^{\sigma y}\mathsf{P}(S_{T-t}^\lambda \geq y)\,dy$$

$$= 2\left(\frac{\sigma + \lambda}{\sigma + 2\lambda}\right)e^{\sigma(\sigma+2\lambda)(T-t)/2}\Phi\left(\frac{-x + (\lambda + \sigma)(T-t)}{\sqrt{T-t}}\right)$$

(2.7)

$$+ e^{\sigma x}\Phi\left(\frac{x - \lambda(T-t)}{\sqrt{T-t}}\right)$$

$$- \left(\frac{\sigma}{\sigma + 2\lambda}\right) e^{(\sigma + 2\lambda)x}\Phi\left(\frac{-x - \lambda(T-t)}{\sqrt{T-t}}\right)$$

*for all $(t,x) \in [0,T] \times \mathbb{R}_+$, and when $\mu = 0$ the function $G$ is given by*

$$G(t,x) = \mathsf{E}(e^{\sigma(x \vee S_{T-t}^{-\sigma/2})}) = e^{\sigma x} + \sigma \int_x^\infty e^{\sigma y}\mathsf{P}(S_{T-t}^{-\sigma/2} \geq y)\,dy$$

(2.8)
$$= \sigma\sqrt{T-t}\,\varphi\left(\frac{x - \sigma(T-t)/2}{\sqrt{T-t}}\right) + e^{\sigma x}\Phi\left(\frac{x + \sigma(T-t)/2}{\sqrt{T-t}}\right)$$

$$+ \left(1 - \sigma x + \frac{\sigma^2}{2}(T-t)\right)\Phi\left(\frac{-x + \sigma(T-t)/2}{\sqrt{T-t}}\right)$$

*for all $(t,x) \in [0,T] \times \mathbb{R}_+$.*

PROOF. Since $B^\lambda$ has stationary independent increments, we see for any (integrable) real-valued $C^1$ function $\Psi$ that

$$\mathsf{E}(\Psi(S_T^\lambda - B_t^\lambda) \mid \mathcal{F}_t^B)$$

(2.9)
$$= \mathsf{E}\left(\Psi\left(\left(S_t^\lambda \vee \max_{t \leq s \leq T} B_s^\lambda\right) - B_t^\lambda\right) \,\Big|\, \mathcal{F}_t^B\right)$$

$$= \mathsf{E}\left(\Psi\left((S_t^\lambda - B_t^\lambda) \vee \max_{0 \leq s \leq T-t}(B_{t+s}^\lambda - B_t^\lambda)\right) \,\Big|\, \mathcal{F}_t^B\right)$$

$$= \mathsf{E}(\Psi(x \vee S_{T-t}^\lambda))|_{x = S_t^\lambda - B_t^\lambda}$$

$$= \left(\Psi(x)\mathsf{P}(S_{T-t}^\lambda \leq x) + \int_x^\infty \Psi(z)\mathsf{P}(S_{T-t}^\lambda \in dz)\right)\bigg|_{x = S_t^\lambda - B_t^\lambda}$$

$$= \Psi(S_t^\lambda - B_t^\lambda) + \int_{S_t^\lambda - B_t^\lambda}^\infty \Psi'(z)\mathsf{P}(S_{T-t}^\lambda > z)\,dz,$$

where the last step follows upon integrating by parts as long as $\lim_{z \to \infty} \Psi(z) \times \mathsf{P}(S_{T-t}^\lambda > z) = 0$.

Turning to (2.4) and setting $\Psi(x) = e^{\sigma x}$ for $x \in \mathbb{R}_+$, we see from (2.5) that $e^{\sigma z}\mathsf{P}(S_{T-t}^\lambda > z) \to 0$ as $z \to \infty$ and therefore

(2.10) $$\mathsf{E}(e^{\sigma(S_T^\lambda - B_t^\lambda)}) = \mathsf{E}(G(t, S_t^\lambda - B_t^\lambda))$$

for all $t \in [0,T]$, where a lengthy calculation based on (2.5) shows that $G$ is given by (2.7) when $\mu \neq 0$ and is given by (2.8) when $\mu = 0$.

Standard arguments based on the fact that each stopping time can be written as the limit of a decreasing sequence of discrete stopping times imply that (2.10) can be extended to

$$\mathsf{E}(e^{\sigma(S_T^\lambda - B_\tau^\lambda)}) = \mathsf{E}(G(\tau, S_\tau^\lambda - B_\tau^\lambda)) \tag{2.11}$$

for all stopping times $\tau$ of $B$ taking values in $[0,T]$, and taking the infimum on both sides over all such stopping times we conclude the proof. □

As it stands, the problem (2.6) appears to be three-dimensional since the underlying Markov process driving the gain function $G$ is the triple $(t, B_t^\lambda, S_t^\lambda)_{0 \le t \le T}$. However, as in [2] and [3], we will show that the problem in fact is only two-dimensional.

Define the process $X = (X_t)_{0 \le t \le T}$ by setting $X_t = S_t^\lambda - B_t^\lambda$ for $t \in [0,T]$. Since $B^\lambda$ is a Lévy process, it follows that $X$ is strong Markov. It is known (cf. [6]) that $X$ is equal in law to $|Y| = (|Y_t|)_{0 \le t \le T}$, where the process $Y = (Y_t)_{0 \le t \le T}$ is the unique strong solution to the stochastic differential equation $dY_t = -\lambda \operatorname{sign}(Y_t) \, dt + dB_t$ with $Y_0 = 0$. It is also known (cf. [6]) that under $Y_0 = x$ the process $|Y|$ has the same law as a Brownian motion with drift $-\lambda$ started at $|x|$ and reflected at 0. Applying the Itô–Tanaka formula to $|Y|$ we see that

$$\begin{aligned} d|Y_t| &= -\lambda \operatorname{sign}(Y_t)^2 \, dt + \operatorname{sign}(Y_t) \, dB_t + d\ell_t^0(Y) \\ &= -\lambda \, dt + d\beta_t + d\ell_t^0(Y), \end{aligned} \tag{2.12}$$

where $\ell^0(Y) = (\ell_t^0(Y))_{0 \le t \le T}$ denotes the local time of $Y$ at zero and $\beta_t = \int_0^t \operatorname{sign}(Y_s) \, dB_s$ defines a standard Brownian motion for $t \in [0,T]$ by Lévy's characterization theorem. Using the equality in law of $X$ and $|Y|$ it follows that the infinitesimal generator $\mathbb{L}_X$ of $X$ acts on functions $f \in C_b^2([0,\infty))$ satisfying $f'(0) = 0$ as

$$\mathbb{L}_X f(x) = -\lambda f'(x) + \tfrac{1}{2} f''(x). \tag{2.13}$$

In order to apply the standard techniques from the theory of optimal stopping for Markov processes (see, e.g., [12]) it is necessary to extend the problem (2.6) by allowing $X$ to start at any time $t \in [0,T]$ at any point $x$ in the state space. It is therefore especially important to see how $X$ depends on its starting value $x$. Although the equation for $Y$ is difficult to solve explicitly, it is known (cf. [2], Lemma 2.2 and [11], Theorem 2.1) that the Markov process $X^x = (X_t^x)_{0 \le t \le T}$ defined under $\mathsf{P}$ as $X_t^x = x \vee S_t^\lambda - B_t^\lambda$ also realizes a Brownian motion with drift $-\lambda$ started at $x \ge 0$ and reflected at 0. Denoting by $\{\mathsf{P}_{t,x} \mid (t,x) \in [0,T] \times \mathbb{R}_+\}$ the family of Markov measures under which $\mathsf{P}_{t,x}(X_t = x) = 1$, it follows that $X$ under $\mathsf{P}_{t,x}$ is equal in law to $X^x$ under $\mathsf{P}$ for any $x \ge 0$ and $t \in [0,T]$ given and fixed.

Using the Markov measures to change the starting point of the process $X$ and letting $\mathsf{E}_{t,x}$ denote expectation under $\mathsf{P}_{t,x}$, we extend the problem (2.6) as follows:

$$V_1(t,x) = \inf_{0 \le \tau \le T-t} \mathsf{E}_{t,x}(G(t+\tau, X_{t+\tau})) = \inf_{0 \le \tau \le T-t} \mathsf{E}(G(t+\tau, X_\tau^x)) \tag{2.14}$$

for any $(t,x) \in [0,T] \times \mathbb{R}_+$. The second equality follows since the infimum in (2.6) is attained at the first entry time $\tau_D$ of $X$ to a closed set $D$ (this follows from general theory of optimal stopping and will be demonstrated below) so that $X_{\tau_D}$ under $\mathsf{P}_{t,x}$ is equally distributed as $X^x_{\tau_D}$ under $\mathsf{P}$. We will freely use either of the representations above without further mention. Note also that $V_1 \leq G$ since one can always insert $\tau \equiv 0$ in (2.14).

**3. The infimum problem.** We are now in a position to prove our main result regarding the infimum problem (2.14). To simplify notation we will write $V$ for the value function $V_1$ from (2.14) throughout this section. We begin by making the following definitions. Define the real-valued function $H$ by

$$
\begin{aligned}
H(t,x) = &\frac{\sigma}{2}(\sigma - 2\lambda)e^{\sigma x}\Phi\left(\frac{x-\lambda(T-t)}{\sqrt{T-t}}\right) \\
&- \frac{\sigma^2}{2}e^{(\sigma+2\lambda)x}\Phi\left(\frac{-x-\lambda(T-t)}{\sqrt{T-t}}\right) \\
&- \sigma(\sigma+\lambda)e^{\sigma(\sigma+2\lambda)(T-t)/2}\Phi\left(\frac{-x+(\sigma+\lambda)(T-t)}{\sqrt{T-t}}\right)
\end{aligned}
\tag{3.1}
$$

for all $(t,x) \in [0,T] \times \mathbb{R}_+$ where $\lambda = (\mu - \sigma^2/2)/\sigma$. Recalling (see, e.g., [7], page 368) that the joint density function of $(B^\lambda_t, S^\lambda_t)$ under $\mathsf{P}$ is given by

$$
f(t,b,s) = \sqrt{\frac{2}{\pi}}\frac{(2s-b)}{t^{3/2}}e^{-(2s-b)^2/(2t)+\lambda(b-\lambda t/2)}
\tag{3.2}
$$

for all $t > 0$, $s \geq 0$ and $b \leq s$, define the functions

$$
\begin{aligned}
J(t,x) &= \mathsf{E}_{t,x}(G(T,X_T)) \\
&= \int_0^\infty \int_{-\infty}^s G(T, x \vee s - b) f(T-t,b,s)\,db\,ds,
\end{aligned}
\tag{3.3}
$$

$$
\begin{aligned}
K(t,x,r,y) &= \mathsf{E}_{t,x}(H(t+r, X_{t+r})I(X_{t+r} > y)) \\
&= \int_0^\infty \int_{-\infty}^s H(t+r, x \vee s - b) \\
&\qquad \times I(x \vee s - b > y)f(r,b,s)\,db\,ds,
\end{aligned}
\tag{3.4}
$$

for all $(t,x) \in [0,T] \times \mathbb{R}_+$, all $r \in [0, T-t]$ and $y \geq 0$. Lastly, the set $\{H \geq 0\} := \{(t,x) \in [0,T] \times \mathbb{R}_+ \mid H(t,x) \geq 0\}$ will play a prominent role in our discussion. A direct examination of the function $H$ reveals the existence of a continuous decreasing function $h \colon [0,T] \to \mathbb{R}_+$ with $h(T) = 0$ such that $\{H \geq 0\} = \{(t,x) \in [0,T] \times \mathbb{R}_+ \mid x \geq h(t)\}$ whenever $\mu \in (0, \sigma^2)$. Our main result in this section may now be stated as follows.

THEOREM 1. *Consider the optimal stopping problem (2.14) and let $D$ denote the optimal stopping set. Then there exists a continuous decreasing function $b:[0,T] \to \mathbb{R}_+$ with $b(T) = 0$ such that*

(3.5) $\quad D = \begin{cases} [0,T] \times \mathbb{R}_+, & \text{when } \mu \leq 0, \\ \{(t,x) \in [0,T] \times \mathbb{R}_+ \mid x \geq b(t)\}, & \text{when } \mu \in (0, \sigma^2), \\ \{(T,x) \mid x \geq 0\}, & \text{when } \mu \geq \sigma^2, \end{cases}$

*so that the stopping time*

(3.6) $\quad\quad\quad \tau_D(t,x) = \inf\{s \in [0, T-t] \mid (t+s, X^x_s) \in D\}$

*is optimal in (2.14) for all $(t,x) \in [0,T] \times \mathbb{R}_+$. More precisely, (3.5) means that when $\mu \leq 0$ it is optimal to stop immediately; when $0 < \mu < \sigma^2$ it is optimal to stop as soon as $X^x$ rises above the curve $b$; and when $\mu \geq \sigma^2$ it is optimal to wait until the final time $T$. Furthermore, the value function from (2.14) is given by*

(3.7) $\quad V(t,x) = \begin{cases} G(t,x), & \text{when } \mu \leq 0, \\ J(t,x) - \int_0^{T-t} K(t,x,s,b(t+s))\,ds, & \text{when } \mu \in (0,\sigma^2), \\ J(t,x), & \text{when } \mu \geq \sigma^2, \end{cases}$

*for all $(t,x) \in [0,T] \times \mathbb{R}_+$, where the function $b$ itself is characterized as the unique solution to the nonlinear Volterra integral equation*

(3.8) $\quad\quad\quad J(t, b(t)) = G(t, b(t)) + \int_0^{T-t} K(t, b(t), s, b(t+s))\,ds$

*in the class of continuous functions $t \mapsto b(t)$ on $[0,T]$ satisfying $b(t) \geq h(t)$ for all $t \in [0,T]$. Finally, the value $V_1$ from (2.4) is given by $V_1 = V(0,0)$ and the optimal stopping time for this problem is $\tau_D(0,0) = \inf\{t \in [0,T] \mid M_t/Z_t \geq e^{\sigma b(t)}\}$.*

PROOF. 1. *Existence of optimal stopping time.* We begin by showing that an optimal stopping time for the problem (2.14) exists. To do this we first establish some general integrability conditions on the function $G$. From the definition of the process $X^x$ we see that

(3.9) $\quad\quad\quad X^x_t = x \vee S^\lambda_t - B^\lambda_t \leq x + 2\lambda T + 2\max_{0 \leq s \leq T}|B_s| =: x + R$

for all $(t,x) \in [0,T] \times \mathbb{R}_+$ where we set $R = 2\lambda T + 2\max_{0 \leq s \leq T}|B_s|$. Turning to the random variable $\max_{0 \leq t \leq T}|B_t|$, observe that $\{\max_{0 \leq t \leq T}|B_t| \geq z\} = \{\max_{0 \leq t \leq T} B_t \geq z\} \cup \{\min_{0 \leq t \leq T} B_t \leq -z\}$ for any $z \geq 0$, so that

(3.10) $\quad \mathsf{P}\left(\max_{0 \leq t \leq T}|B_t| \geq z\right) \leq \mathsf{P}\left(\max_{0 \leq t \leq T} B_t \geq z\right) + \mathsf{P}\left(\max_{0 \leq t \leq T}(-B_t) \geq z\right)$

$\quad\quad\quad\quad\quad = 2\mathsf{P}\left(\max_{0 \leq t \leq T} B_t \geq z\right) = 2\mathsf{P}(|B_T| \geq z)$

since the random variables $\max_{0 \leq t \leq T} B_t$ and $|B_T|$ are equal in law. A similar calculation as at (2.9) then shows that

$$
\begin{aligned}
\mathsf{E}(e^{\alpha R}) &= \mathsf{E}(e^{\alpha(2\lambda T + \max_{0 \leq t \leq T} |B_t|)}) \\
&= e^{2\alpha\lambda T}\left(1 + \int_0^\infty \alpha e^{\alpha z} \mathsf{P}\left(\max_{0 \leq t \leq T} |B_t| \geq z\right) dz\right) \\
&\leq e^{2\alpha\lambda T}\left(1 + \int_0^\infty 2\alpha e^{\alpha z} \mathsf{P}(|B_T| \geq z) \, dz\right) < \infty
\end{aligned}
\tag{3.11}
$$

for any $\alpha \in \mathbb{R}$. Turning to (2.7) and (2.8) and using (3.9) above, we see that

$$
0 \leq G(t, X_t^x) \leq K_1 + K_2 e^{\sigma(x+R)} + K_3 e^{(1+|\sigma+2\lambda|)(x+R)}, \tag{3.12}
$$

where $K_1$, $K_2$ and $K_3$ are positive constants (independent of $t$). This combined with (3.11) shows that $G(t, X_t^x)$ is bounded by an integrable random variable for all $t \in [0, T]$ and $x \geq 0$.

Using the dominated convergence theorem together with the continuity of the function $G$ and the continuity of the flow $x \mapsto X^x$, we see that the map $(t, x) \mapsto \mathsf{E}(G(t + \tau, X_\tau^x))$ is continuous and thus upper semicontinuous (usc) for every stopping time $\tau$ taking values in $[0, T-t]$. Since the infimum of usc functions is usc, it follows that the function $V$ is usc as well and so by general results of optimal stopping (see [12], Corollary 2.9) we conclude that an optimal stopping time for the problem (2.14) exists. Moreover, this stopping time is given by (3.6) above where the stopping set is given by $D = \{(t, x) \in [0, T] \times \mathbb{R}_+ \mid V(t, x) = G(t, x)\}$ and the continuation set $C$ is given by $C = D^c = \{(t, x) \in [0, T] \times \mathbb{R}_+ \mid V(t, x) < G(t, x)\}$. The fact that $D$ is closed (and $C$ is open) follows from the fact that $V$ is usc.

2. *Shape of $D$.* We now turn to the question of determining the shape of the stopping set $D$. From either (2.7) or (2.8) above, note that

$$
G_x(t, x) = \sigma e^{\sigma x} \mathsf{P}(S_{T-t}^\lambda \leq x) \leq \sigma e^{\sigma x} \tag{3.13}
$$

so that in particular $G_x(t, 0) = 0$ for all $t \in [0, T)$. By Itô's formula we get

$$
\begin{aligned}
G(t+s, X_s^x) &= G(t, x) + \int_0^s \left(G_t - \lambda G_x + \frac{1}{2} G_{xx}\right)(t+r, X_r^x) \, dr \\
&\quad + \int_0^s G_x(t+r, X_r^x) \, d(x \vee S_r^\lambda) - \int_0^s G_x(t+r, X_r^x) \, dB_r \\
&= G(t, x) + \int_0^s H(t+r, X_r^x) \, dr + M_s,
\end{aligned}
\tag{3.14}
$$

where we use that $d(x \vee S_r^\lambda)$ is zero off the set of all $r \in [0, s]$ at which $X_r^x \neq 0$ while $G_x(t+r, X_r^x) = 0$ for $X_r^x = 0$, and we set $M_s = -\int_0^s G_x(t+r, X_r^x) \, dB_r$ for $s \in [0, T-t]$. A lengthy calculation shows that the function $H = G_t - \lambda G_x + \frac{1}{2} G_{xx}$ is given by (3.1) above when $\lambda \neq -\frac{\sigma}{2}$, and is given by

$$(3.15) \quad H(t,x) = \sigma^2 e^{\sigma x} \Phi\left(\frac{x + \sigma(T-t)/2}{\sqrt{T-t}}\right) - \sigma^2 \Phi\left(\frac{-x + \sigma(T-t)/2}{\sqrt{T-t}}\right)$$

for all $(t,x) \in [0,T] \times \mathbb{R}_+$ when $\lambda = -\frac{\sigma}{2}$. Equations (3.9), (3.11) and (3.13) together with the Burkholder–Davis–Gundy inequalities show that the local martingale $M = (M_s)_{s \in [0,T-t]}$ in (3.14) is a martingale. Replacing $s$ in (3.14) with $\tau_D(t,x)$, taking expectations and using the optional sampling theorem, we obtain

$$(3.16) \quad V(t,x) = G(t,x) + \mathsf{E}\left(\int_0^{\tau_D(t,x)} H(t+r, X_r^x)\, dr\right)$$

for all $(t,x) \in [0,T] \times \mathbb{R}_+$.

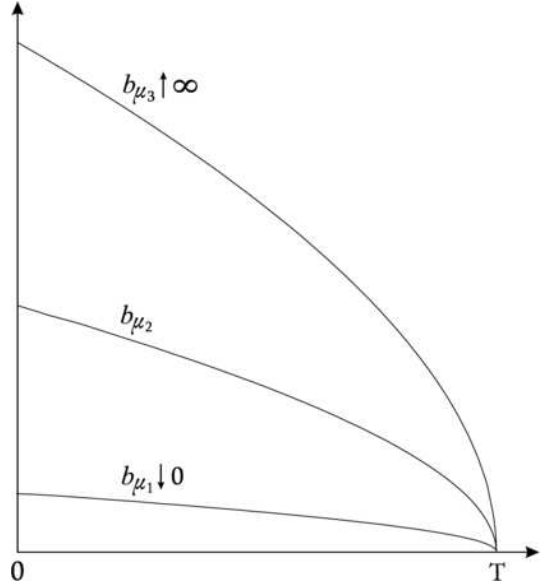

FIG. 1. *The optimal stopping boundaries in the optimal prediction problem (2.4) for drifts $\mu_1$, $\mu_2$ and $\mu_3$ satisfying $0 < \mu_1 < \mu_2 < \mu_3 < \sigma^2$. The optimal stopping sets lie above the boundaries. The convergence relations take place for $\mu_1 \downarrow 0$ and $\mu_3 \uparrow \sigma^2$.*

Recall that $\lambda = (\mu - \sigma^2/2)/\sigma$ and let us suppose first that $\lambda = -\frac{\sigma}{2}$ (i.e., $\mu = 0$). Note from (3.15) that

$$(3.17) \quad H(t,x) \geq \sigma^2 \left[\Phi\left(\frac{x + \sigma(T-t)/2}{\sqrt{T-t}}\right) - \Phi\left(\frac{-x + \sigma(T-t)/2}{\sqrt{T-t}}\right)\right]$$
$$> 0$$

for all $(t, x) \in [0, T) \times (0, \infty)$. Choosing any $t \in [0, T)$ and $x \geq 0$, we see from (3.16) that we must have $\tau_D(t, x) = 0$ since otherwise we would have $V(t, x) > G(t, x)$ which is a contradiction. Therefore when $\mu = 0$ we see that $\tau_D \equiv 0$ so that the optimal stopping set $D$ is given by $D = [0, T] \times \mathbb{R}_+$.

A similar result holds when we assume that $\lambda < -\frac{\sigma}{2}$ (i.e., $\mu < 0$). Turning to (3.1), we see that

$$\begin{aligned}
(3.18) \quad H(t, x) &\geq \sigma^2 e^{\sigma x} \Phi\left(\frac{x - \lambda(T - t)}{\sqrt{T - t}}\right) - \frac{\sigma^2}{2} e^{(\sigma + 2\lambda)x} \Phi\left(\frac{-x - \lambda(T - t)}{\sqrt{T - t}}\right) \\
&\quad - \frac{\sigma^2}{2} e^{\sigma(\sigma + 2\lambda)(T - t)/2} \Phi\left(\frac{-x + (\sigma + \lambda)(T - t)}{\sqrt{T - t}}\right) \\
&\geq \frac{\sigma^2}{2}\left[\Phi\left(\frac{x - \lambda(T - t)}{\sqrt{T - t}}\right) - \Phi\left(\frac{-x - \lambda(T - t)}{\sqrt{T - t}}\right)\right] \\
&\quad + \frac{\sigma^2}{2}\left[\Phi\left(\frac{x + \sigma(T - t)/2}{\sqrt{T - t}}\right) - \Phi\left(\frac{-x + \sigma(T - t)/2}{\sqrt{T - t}}\right)\right] > 0
\end{aligned}$$

for all $(t, x) \in [0, T) \times (0, \infty)$ where the inequalities follow since $\sigma - 2\lambda > 2\sigma$ while $\sigma + \lambda < \sigma/2$ and $\sigma + 2\lambda < 0$. Turning to (3.16) and choosing any $t \in [0, T)$ and $x \geq 0$, we see that we must have $\tau_D(t, x) = 0$ since otherwise we would have $V(t, x) > G(t, x)$.

We conclude therefore that whenever $\mu \leq 0$, we have $D = [0, T] \times \mathbb{R}_+$ and $\tau_D \equiv 0$ so that it is optimal to stop immediately. This establishes the first parts of (3.5) and (3.7) in Theorem 1 above.

Suppose now that $\lambda \geq \frac{\sigma}{2}$ (i.e., $\mu \geq \sigma^2$) so that $\sigma - 2\lambda \leq 0$. From (3.1) above we easily see that $H(t, x) < 0$ for all $(t, x) \in [0, T] \times [0, \infty)$. Choose now any $t \in [0, T)$ and $x \geq 0$, let $U \subseteq [0, T) \times \mathbb{R}_+$ be an open neighborhood of $(t, x)$ and denote by $\sigma_U$ the first exit time from $U$ when $X$ starts at $x$ at time $t$. Replacing $s$ with $\sigma_U$ in (3.14) above, taking expectations and using the optional sampling theorem, we see that

$$\begin{aligned}
(3.19) \quad V(t, x) &\leq \mathsf{E}(G(t + \sigma_U, X^x_{\sigma_U})) = G(t, x) + \mathsf{E}\left(\int_0^{\sigma_U} H(t + s, X^x_s)\, ds\right) \\
&< G(t, x),
\end{aligned}$$

which shows that $(t, x) \in C$. Therefore all points $(t, x) \in [0, T) \times \mathbb{R}_+$ must lie in the continuation set, so that it is never optimal to stop before the end of time. We see then that when $\mu \geq \sigma^2$, we have $\tau_D(t, x) = T - t$ for all $(t, x) \in [0, T] \times \mathbb{R}_+$ so that $D = \{(T, x) \mid x \geq 0\}$. This establishes the last part of (3.5) above, and since $V(t, x) = \mathsf{E}_{t,x}(G(t + \tau_D, X_{t + \tau_D})) = \mathsf{E}_{t,x}(G(T, X_T)) = J(t, x)$ from (3.3), the last part of (3.7) holds as well.

To summarize, we have shown that when $\mu \leq 0$ or $\mu \geq \sigma^2$ the optimal stopping problem (2.14) has a trivial solution: in the first case it is always optimal to stop immediately, whereas in the last case it is always optimal to wait until the end of time. Our task therefore reduces to describing the solution of (2.14) when $\mu$ does not lie in either of these sets. In the remainder of the proof we will therefore assume that $\mu \in (0, \sigma^2)$, that is, $\lambda \in (-\sigma/2, \sigma/2)$.

When $\lambda$ is constrained to this interval, a direct examination of $H$ from (3.1) reveals the existence of a continuous decreasing function $h$ on $[0,T]$ with $h(T) = 0$ such that $\{H < 0\} = \{(t,x) \in [0,T] \times \mathbb{R}_+ \mid x < h(t)\}$. Arguments similar to (3.19) above then show that $\{H < 0\} \subseteq C$, and defining the optimal stopping boundary $b$ as

$$(3.20) \qquad b(t) = \inf\{x \geq 0 \mid (t,x) \in D\}$$

for all $t \in [0,T]$ it follows that $b(t) \geq h(t)$ for all $t \in [0,T]$. Moreover, differentiating (3.1) in time we find that

$$\begin{aligned}
H_t(t,x) &= \frac{\sigma^2}{2} e^{\sigma x} \left( \frac{2x + (\sigma + 2\lambda)(T-t)}{(T-t)^{3/2}} \right) \varphi\left( \frac{x - \lambda(T-t)}{\sqrt{T-t}} \right) \\
(3.21) \qquad &\quad + \frac{\sigma^2}{2}(\sigma + \lambda)(\sigma + 2\lambda) e^{\sigma(\sigma+2\lambda)(T-t)/2} \\
&\quad \times \Phi\left( \frac{-x + (\sigma+\lambda)(T-t)}{\sqrt{T-t}} \right) \geq 0
\end{aligned}$$

for all $(t,x) \in [0,T] \times \mathbb{R}_+$ whenever $\lambda \in (-\sigma/2, \sigma/2)$. To see the importance of this fact, fix any $x \in \mathbb{R}_+$ and $s < t$ in $[0,T]$ and set $\tau_s = \tau_D(s,x)$ and $\tau_t = \tau_D(t,x)$. Since $0 \leq \tau_t \leq T-t < T-s$ and since $\tau_t$ is a suboptimal time when $X$ starts at $x$ at time $s$, we see from (3.16) that

$$\begin{aligned}
&(V(t,x) - G(t,x)) - (V(s,x) - G(s,x)) \\
(3.22) \qquad &\geq \mathsf{E}\left( \int_0^{\tau_t} H(t+r, X_r^x)\,dr \right) - \mathsf{E}\left( \int_0^{\tau_t} H(s+r, X_r^x)\,dr \right) \\
&= \mathsf{E}\left( \int_0^{\tau_t} H(t+r, X_r^x) - H(s+r, X_r^x)\,dr \right) \geq 0,
\end{aligned}$$

from where we derive the important fact that

$$(3.23) \qquad t \mapsto V(t,x) - G(t,x) \text{ is increasing on } [0,T]$$

for each $x \in \mathbb{R}_+$ given and fixed. A direct consequence of this is that if any point $(t,x) \in D$, then all points $(t+s, x) \in D$ for $s \in [0, T-t]$ since $0 \geq V(t+s, x) - G(t+s, x) \geq V(t,x) - G(t,x) = 0$. This means that the function $t \mapsto b(t)$ is decreasing.

We now show that if $(t,x) \in D$, then all points $(t,y) \in D$ for $y \geq x$. To see this, fix a point $(t,x) \in D$ and take any $y \geq x$. Since all the points $(t+s, x) \in D$ for $s \in [0, T-t]$, the process $X$ started at $(t,y)$ must enter the stopping set $D$ upon (or before) hitting the level $x$. In particular, we must have $X_{t+s} \geq x$ for all $s \in [0, \tau_D(t,y)]$ under $\mathsf{P}_{t,y}$. However, recalling that $x \geq h(t)$ and that $h$ is decreasing, it follows that the rectangle $[t,T] \times [x,\infty) \subseteq \{H \geq 0\}$ and so from (3.16) we must have $V(t,y) \geq G(t,y)$. This shows that $(t,y) \in D$ for all $y \geq x$ whenever $(t,x) \in D$, and combined with the observations above establishes that $D$ has the form given in (3.5) above.

3. *Function $b$ is finite-valued.* We show that the optimal stopping boundary $b$ is finite valued, which also shows that the stopping set $D$ is strictly greater than the set $\{(T,x) \mid x \in \mathbb{R}_+\}$. Suppose that the function $b(t)$ is not finite-valued for all $t \in [0,T]$ and define the time $t_* \in [0,T]$ as $t_* = \sup\{t \in [0,T] \mid b(t) = \infty\}$. Consider first the case when $t_* \in (0,T]$ and note that there are two possibilities: either $b$ has a jump discontinuity at $t_*$ jumping down from infinity to a finite value, or $b$ has an asymptote at $t_*$. Setting $\tau_x = \tau_D(0,x)$, it is clear that $\tau_x \to t_*$ as $x \to \infty$ in either case. From (3.1) we see that $m := \inf\{H(t,x) \mid (t,x) \in [0,T] \times \mathbb{R}_+\} > -\infty$, and so by Fatou's lemma and (3.16) it follows that

$$
\begin{aligned}
0 &\geq \liminf_{x \to \infty}(V(0,x) - G(0,x)) \\
&= \liminf_{x \to \infty}\left[\mathsf{E}\left(\int_0^{\tau_x} H(t, X_t^x)I(X_t^x > h(t))\,dt\right) \right. \\
&\qquad\qquad\left. + \mathsf{E}\left(\int_0^{\tau_x} H(t, X_t^x)I(X_t^x \leq h(t))\,dt\right)\right] \\
&\geq \mathsf{E}\left(\int_0^{t_*} H(t,\infty)\,dt\right) + mT = \infty,
\end{aligned}
$$
(3.24)

which shows that we cannot have $t_* \in (0,T]$. On the other hand if $t_* = 0$, then by extending the terminal time to $T' > T$ and considering our problem on the interval $[0,T']$ instead of $[0,T]$ we will make $t_*$ strictly positive (since $b$ is decreasing), reducing it to the case already considered. Therefore $b(t)$ must be finite for all $t \in [0,T]$.

4. *Continuity of $V$.* We show that $(t,x) \mapsto V(t,x)$ is continuous on $[0,T] \times \mathbb{R}_+$. For this, take any $t \in [0,T]$ and $x,y \in \mathbb{R}_+$, set $\tau_x = \tau_D(t,x)$ and $\tau_y = \tau_D(t,y)$ and suppose without loss of generality that $x \leq y$. From (2.14) we see that

$$
\begin{aligned}
\mathsf{E}(G(t+\tau_y, X_{\tau_y}^y) &- G(t+\tau_y, X_{\tau_y}^x)) \\
&\leq V(t,y) - V(t,x) \\
&\leq \mathsf{E}(G(t+\tau_x, X_{\tau_x}^y) - G(t+\tau_x, X_{\tau_x}^x)).
\end{aligned}
$$
(3.25)

Note from (3.13) that $x \mapsto G(t,x)$ is increasing. This together with the mean value theorem and (3.9) shows that for any stopping time $\tau \in [0, T-t]$ we have

$$0 \leq G(t+\tau, X_\tau^y) - G(t+\tau, X_\tau^x) = G_x(t+\tau, \xi)(X_\tau^y - X_\tau^x)$$

(3.26)
$$\leq \sigma e^{\sigma \xi}(X_\tau^y - X_\tau^x) \leq \sigma e^{\sigma(y+R)}(y-x),$$

where $\xi$ is some value in $[X_\tau^x, X_\tau^y]$, and the last inequality is obvious once we observe that $X_\tau^y - X_\tau^x = y \vee S_\tau^\lambda - x \vee S_\tau^\lambda \leq y - x$ and $\xi \leq X_\tau^y \leq y + R$. Inserting (3.26) in (3.25) yields

(3.27) $$0 \leq V(t,y) - V(t,x) \leq \sigma \mathsf{E}(e^{\sigma(y+R)})(y-x)$$

and taking the limit as $y - x \to 0$ we see that $x \mapsto V(t,x)$ is continuous on $\mathbb{R}_+$ uniformly over all $t \in [0,T]$.

To complete the proof of the initial claim it is enough to show that $t \mapsto V(t,x)$ is continuous on $[0,T]$ for every $x \in \mathbb{R}_+$. For this fix $x \geq 0$, take any $s \leq t$ in $[0,T]$ and set $\tau_s = \tau_D(s,x)$. Since the stopping time $\tau_s$ does not necessarily lie in the interval $[0, T-t]$, it is not possible to mimic the previous argument directly and we therefore adjust our approach as follows. Recalling (3.23) and defining the stopping time $\rho = \tau_s \wedge (T-t)$, we see from (3.14) and (3.16) upon using the optional sampling theorem that

(3.28)
$$\begin{aligned}
0 &\leq V(t,x) - G(t,x) - (V(s,x) - G(s,x)) \\
&\leq \mathsf{E}\left(\int_0^\rho H(t+r, X_r^x) - H(s+r, X_r^x)\,dr\right) \\
&\quad - \mathsf{E}\left(\int_\rho^{\tau_s} H(s+r, X_r^x)\,dr\right).
\end{aligned}$$

Note from (3.1) that

(3.29) $$|H(t,x)| \leq \frac{\sigma}{2}(\sigma - 2\lambda)e^{\sigma x} + \frac{\sigma^2}{2}e^{(\sigma+2\lambda)x} + \sigma(\sigma+\lambda)e^{\sigma(\sigma+2\lambda)T/2}$$

for all $(t,x) \in [0,T] \times [0,\infty)$, and since $0 \leq \tau_s - \rho \leq t - s$ we may pass to the limit as $t - s \to 0$ in (3.28) and use the dominated convergence theorem [recalling (3.9) and (3.11) for the necessary integrability] to conclude that $t \mapsto V(t,x) - G(t,x)$ is continuous. The continuity of $t \mapsto G(t,x)$ then completes the proof.

5. *Free-boundary problem.* We now formulate a free-boundary problem that the value function $V$ solves. This differential equation will be useful to us later on, but is also interesting in its own right and can be used as the departure point in computing numerical values for the optimal stopping boundary $b$ and for the value function $V$. It is well known from the theory of Markov processes (see, e.g., [12], Chapter III, Section 7) that $V$ is $C^{1,2}$ in the continuation set $C$ and satisfies the following version of the *Kolmogorov backward equation*:

(3.30) $$V_t(t,x) - \lambda V_x(t,x) + \tfrac{1}{2}V_{xx}(t,x) = 0 \qquad \text{for all } (t,x) \in C$$

together with the following instantaneous stopping condition:

(3.31) $$V(t,x) = G(t,x) \quad \text{for all } (t,x) \in D.$$

The free-boundary problem is completed by the *normal reflection* condition and the *smooth fit* condition respectively:

(3.32) $\quad V_x(t,0+) = 0 \quad$ for all $t \in [0,T)$,

(3.33) $\quad x \mapsto V_x(t,x)$ is continuous at $b(t) \quad$ for all $t \in [0,T)$,

which will be established below. The smooth fit condition in particular will play an important role in the derivation of the integral equations (3.7) and (3.8).

6. *Smooth fit.* We show that $x \mapsto V_x(t,x)$ is continuous over the optimal stopping boundary $b$. Fix any $t \in [0,T)$ and $\varepsilon > 0$ and set $x = b(t)$ and $\tau_\varepsilon = \tau_D(t, x - \varepsilon)$. We first show that $V_x(t,x) = G_x(t,x)$. From the mean value theorem we have

(3.34)
$$\begin{aligned}
G(t,x) - G(t, x-\varepsilon) &\leq V(t,x) - V(t, x-\varepsilon) \\
&\leq \mathsf{E}(G(t+\tau_\varepsilon, X^x_{\tau_\varepsilon}) - G(t+\tau_\varepsilon, X^{x-\varepsilon}_{\tau_\varepsilon})) \\
&= \mathsf{E}(G_x(t+\tau_\varepsilon, \xi_\varepsilon)(X^x_{\tau_\varepsilon} - X^{x-\varepsilon}_{\tau_\varepsilon})) \\
&\leq \varepsilon \mathsf{E}(G_x(t+\tau_\varepsilon, \xi_\varepsilon)),
\end{aligned}$$

where $X^{x-\varepsilon}_{\tau_\varepsilon} \leq \xi_\varepsilon \leq X^x_{\tau_\varepsilon}$. Since the optimal stopping boundary $b$ is decreasing it follows that all points $(t+s, x) \in D$ for $s \in [0, T-t]$. Define the stopping time

(3.35) $$\sigma_\varepsilon = \inf\{s \in [0, T-t] \mid X^{x-\varepsilon}_s \geq x\}$$

and note that $0 \leq \tau_\varepsilon \leq \sigma_\varepsilon$. Then under $\mathsf{P}$ we have

(3.36)
$$\begin{aligned}
\sigma_\varepsilon &= \inf\{s \in [0, T-t] \mid (x-\varepsilon) \vee S^\lambda_s - B^\lambda_s \geq x\} \\
&\leq \inf\{s \geq 0 \mid -B_s \geq \varepsilon + \lambda s\} \to 0
\end{aligned}$$

as $\varepsilon \to 0$ since the function $t \mapsto \varepsilon + \lambda t$ is a lower function of Brownian motion at $0+$. It follows therefore that $\tau_\varepsilon \to 0$ as $\varepsilon \downarrow 0$ as well.

Turning to (3.34), dividing through by $\varepsilon$ and passing to the limit as $\varepsilon \downarrow 0$, we see that the first term converges to $G_x(t,x)$ since $G$ is differentiable, while the last term converges to $\mathsf{E}(G_x(t, X^x_0)) = G_x(t,x)$ by dominated convergence [upon recalling (3.11) and (3.13)] since $\xi_\varepsilon \to X^x_0$. We conclude therefore that $V_x(t,x) = G_x(t,x)$ as claimed.

A small modification of the argument above shows that $x \mapsto V(t,x)$ is continuously differentiable at $b(t)$. Indeed, taking $\delta > 0$ and setting $\tau_\delta =$

$\tau_D(t, x-\delta)$ with $x = b(t)$, we see as before that for any $\varepsilon \in (0, \delta)$ we have

$$V(t, x-\delta+\varepsilon) - V(t, x-\delta)$$
(3.37)
$$\leq \mathsf{E}(G(t+\tau_\delta, X^{x-\delta+\varepsilon}_{\tau_\delta}) - G(t+\tau_\delta, X^{x-\delta}_{\tau_\delta}))$$
$$= \mathsf{E}(G_x(t+\tau_\delta, \eta)(X^{x-\delta+\varepsilon}_{\tau_\delta} - X^{x-\delta}_{\tau_\delta})) \leq \varepsilon \mathsf{E}(G_x(t+\tau_\delta, \eta)),$$

where $\eta \in [X^{x-\delta}_{\tau_\delta}, X^{x-\delta+\varepsilon}_{\tau_\delta}]$. Clearly $\eta \to X^{x-\delta}_{\tau_\delta}$ as $\varepsilon \to 0$, and in a similar manner to (3.35) and (3.36) above we can show that $\tau_\delta \to 0$ as $\delta \to 0$. Dividing (3.37) by $\varepsilon$ and taking first the limit as $\varepsilon \downarrow 0$ (recalling that $V$ is $C^{1,2}$ in $C$ so that $V_x$ exists) and then the limsup as $\delta \downarrow 0$, we see by the dominated convergence theorem that

(3.38) $$\limsup_{\delta \downarrow 0} V_x(t, x-\delta) \leq G_x(t, x).$$

To prove the reverse inequality, take $\varepsilon > 0$ and note that

$$\frac{V(t, x-\delta) - V(t, x-\delta-\varepsilon)}{\varepsilon}$$
(3.39)
$$\geq \frac{1}{\varepsilon}\mathsf{E}(G(t+\tau_\delta, X^{x-\delta}_{\tau_\delta}) - G(t+\tau_\delta, X^{x-\delta-\varepsilon}_{\tau_\delta}))$$
$$= \mathsf{E}\left(\frac{1}{\varepsilon}(X^{x-\delta}_{\tau_\delta} - X^{x-\delta-\varepsilon}_{\tau_\delta})G_x(t+\tau_\delta, \eta)\right),$$

where $\eta \in [X^{x-\delta-\varepsilon}_{\tau_\delta}, X^{x-\delta}_{\tau_\delta}]$. Observe that $0 \leq \frac{1}{\varepsilon}(X^{x-\delta}_{\tau_\delta} - X^{x-\delta-\varepsilon}_{\tau_\delta}) \leq 1$ and

(3.40) $$\frac{1}{\varepsilon}(X^{x-\delta}_{\tau_\delta} - X^{x-\delta-\varepsilon}_{\tau_\delta}) \to I(S^\lambda_{\tau_\delta} < x - \delta)$$

as $\varepsilon \to 0$, while clearly $\eta \to X^{x-\delta}_{\tau_\delta}$. Passing to the limit as $\varepsilon \downarrow 0$ in (3.39) and using the dominated convergence theorem, we see that

(3.41) $$V_x(t, x-\delta) \geq \mathsf{E}(I(S^\lambda_{\tau_\delta} < x-\delta)G_x(t+\tau_\delta, X^{x-\delta}_{\tau_\delta})).$$

Taking the liminf as $\delta \downarrow 0$ and recalling that $\tau_\delta \to 0$, we obtain the reverse inequality

(3.42) $$\liminf_{\delta \downarrow 0} V_x(t, x-\delta) \geq G_x(t, x)$$

and conclude the result.

7. *Continuity of $b$.* We show that the function $t \mapsto b(t)$ is continuous on $[0, T]$. We begin by proving that $b$ is right-continuous. For this fix any $t \in [0, T)$, let $t_n \downarrow t$ and consider the limit $b(t+) := \lim_{n \to \infty} b(t_n)$ which exists as $b$ is decreasing. Since $(t_n, b(t_n)) \in D$ for all $n \geq 1$ and $D$ is closed, it follows that $(t, b(t+)) \in D$ and so from (3.20) we see that $b(t) \leq b(t+)$. On the other hand, the fact that $b$ is decreasing implies that $b(t) \geq b(t_n)$ for all $n \geq 1$, and passing to the limit as $n \to \infty$ we obtain the reverse inequality.

We now show that $b$ is left-continuous. Suppose this is not the case so that there exists some $t \in (0,T]$ at which $b(t-) > b(t)$, and choose any $x \in (b(t), b(t-))$. Since $b \geq h$ and $h$ is continuous, it follows that $x > h(s)$ for all $s \in [s_1, t]$ for some $s_1$ sufficiently close to $t$. Hence $m := \inf\{H(s,y) \mid s \in [s_1, t), y \in [x, b(s)]\} > 0$ by the continuity of $H$. Moreover, since $V$ is continuous and $V(t,y) = G(t,y)$ for all $y \in [x, b(t-)]$, it follows that

$$(3.43) \qquad |\lambda(V(s,y) - G(s,y))| \leq \frac{m}{4}(b(t-) - x)$$

for all $s \in [s_2, t]$ and $y \in [x, b(s)]$ where $s_2 \in [s_1, t)$ is some value sufficiently close to $t$. Since $H = G_t - \lambda G_x + \frac{1}{2}G_{xx}$ we see from (3.23) and (3.30) that $\frac{1}{2}(V_{xx} - G_{xx}) = G_t - V_t + \lambda(V_x - G_x) - H \leq \lambda(V_x - G_x) - H$, and this together with the smooth-fit condition (3.33) and (3.43) implies that

$$(3.44) \quad \begin{aligned} V(s,x) - G(s,x) &= \int_x^{b(s)} \int_y^{b(s)} (V_{xx}(s,z) - G_{xx}(s,z))\,dz\,dy \\ &\leq 2\int_x^{b(s)} \int_y^{b(s)} (\lambda(V_x - G_x) - H)(s,z)\,dz\,dy \\ &\leq 2\int_x^{b(s)} (-\lambda(V(s,y) - G(s,y)) - m(b(s)-y))\,dy \\ &\leq \frac{m}{2}(b(t-) - x)(b(s) - x) - m(b(s) - x)^2 \end{aligned}$$

for any $s \in [s_2, t)$. Passing to the limit as $s \uparrow t$ gives $V(t,x) - G(t,x) \leq -\frac{m}{2}(b(t-) - x)^2 < 0$ and contradicts the fact that $(t,x) \in D$. We conclude therefore that $t \mapsto b(t)$ is continuous on $[0,T]$. Note that this proof also shows that $b(T) = 0$ since $h(T) = 0$ and $V(T,x) = G(T,x)$ for all $x \in \mathbb{R}_+$.

8. *Normal reflection.* We show that the normal reflection condition (3.32) holds. For this, first note from (3.27) that $x \mapsto V(t,x)$ is increasing on $[0,\infty)$ so that $V_x(t,0+) \geq 0$ for all $t \in [0,T)$. Note that the limit exists since $V$ is $C^{1,2}$ in $C$. Suppose that there exists some $t \in [0,T)$ such that $V_x(t,0+) > 0$. The smoothness of $V$ in $C$ implies that $t \mapsto V_x(t,0+)$ is continuous on $[0,T)$, and so there must exist a $\delta > 0$ such that $V_x(t+s, 0+) \geq \varepsilon > 0$ for all $s \in [0,\delta]$ with $t + \delta < T$. Setting $\tau_\delta = \tau_D \wedge \delta$, recalling (3.30) and applying Itô's formula to $V$ in $C$, we see that

$$(3.45) \quad \begin{aligned} \mathsf{E}(V(t+\tau_\delta, X_{\tau_\delta}^0)) &= V(t,0) + \mathsf{E}\left(\int_0^{\tau_\delta} V_x(t+s, X_s^0)\,dS_s^\lambda\right) \\ &\geq V(t,0) + \varepsilon \mathsf{E}(S_{\tau_\delta}^\lambda) \end{aligned}$$

by the optional sampling theorem, where the latter follows from (3.11) upon dividing (3.27) by $y - x$ and passing to the limit as $y - x \to 0$. From the general theory of optimal stopping for Markov processes (see, e.g., [12]) we know that the process $(V(t+s \wedge \tau_D, X^0_{s \wedge \tau_D}))_{0 \leq s \leq T-t}$ is a martingale. This means that we must have $\mathsf{E}(S^\lambda_{\tau_\delta}) = 0$, but since the properties of the process $S^\lambda$ clearly exclude this, we conclude that $V_x(t, 0+) = 0$ for all $t \in [0, T)$.

9. *Integral equations.* We may now derive the integral equations (3.7) and (3.8). Setting $c = b(0) + 1$ we see from (3.27) that

$$(3.46) \qquad |\lambda| V_x(t, x) \leq |\lambda| \sigma \mathsf{E}(e^{\sigma(c+R)}) =: K < \infty$$

for all $(t, x) \in [0, T] \times [0, c]$. Using this inequality in (3.30) and recalling (3.23) we obtain $\frac{1}{2} V_{xx} = -V_t + \lambda V_x \leq -G_t + K$ in $C$. If we set

$$(3.47) \qquad f(t, x) = 2 \int_0^x \int_0^y (-G_t(t, z) + K) \, dz \, dy$$

for all $(t, x) \in [0, T] \times [0, c]$, we see that $V_{xx} \leq f_{xx}$ on $[0, T] \times [0, c]$. Defining the function $F : [0, T] \times \mathbb{R}_+ \to \mathbb{R}$ by $F(t, x) = V(t, x) - f(t, x)$, we see that: (i) the map $x \mapsto F(t, x)$ is concave on the intervals $[0, b(t))$ and $(b(t), c]$ for every $t \in [0, T]$; (ii) the function $F$ is $C^{1,2}$ on $C \cup D^o$; (iii) the function $F_t - \lambda F_x + \frac{1}{2} F_{xx}$ is locally bounded on $C \cup D^o$; and (iv) the map $t \mapsto F_x(t, b(t) \pm) = G_x(t, b(t)) - f_x(t, b(t))$ is continuous on $[0, T]$. Since the function $b$ is decreasing and consequently of bounded variation, we may apply the local time–space formula [10] to $F(t+s, X_{t+s})$ and Itô's formula to $f(t+s, X_{t+s})$ since $f$ is $C^{1,2}$. Adding these two expressions, using (3.30), (3.32), (3.33) and the fact that $f_x(t, 0) = 0$, we obtain

$$
\begin{aligned}
V(t+s, X^x_s) \\
&= V(t, x) + \int_0^s \left( V_t - \lambda V_x + \frac{1}{2} V_{xx} \right)(t+r, X^x_r) I(X^x_r \neq b(t+r)) \, dr \\
&\quad + \int_0^s V_x(t+r, X^x_r) I(X^x_r \neq b(t+r)) \, d(x \vee S^\lambda_r) \\
&\quad - \int_0^s V_x(t+r, X^x_r) I(X^x_r \neq b(t+r)) \, dB_r \\
&\quad + \frac{1}{2} \int_0^s (V_x(t+r, X^x_r+) - V_x(t+r, X^x_r-)) \\
&\qquad \times I(X^x_r = b(t+r)) \, d\ell^b_r(X^x) \\
&= V(t, x) + \int_0^s H(t+r, X^x_r) I(X^x_r > b(t+r)) \, dr + M_s
\end{aligned}
$$
(3.48)

for any $(t, x) \in [0, T] \times \mathbb{R}_+$ and $s \in [0, T-t]$, where $\ell^b(X^x)$ denotes the local time of $X^x$ on the curve $b$ and $M_s = -\int_0^s V_x(t+r, X^x_r) \, dB_r$ is a martingale for $s \in [0, T-t]$. Setting $s = T - t$, taking expectations and using the optional sampling theorem, we obtain

$$V(t, x) = \mathsf{E}_{t,x}(G(T, X_T))$$

(3.49)
$$- \mathsf{E}_{t,x}\left(\int_0^{T-t} H(t+r, X_{t+r}) I(X_{t+r} > b(t+r))\, dr\right)$$

which is exactly (3.7) after interchanging the order of integration. Setting $x$ equal to $b(t)$ in (3.49) we get

(3.50)
$$G(t, b(t)) = \mathsf{E}_{t,b(t)}(G(T, X_T))$$
$$- \int_0^{T-t} \mathsf{E}_{t,b(t)}(H(t+r, X_{t+r}) I(X_{t+r} > b(t+r)))\, dr$$

which is exactly (3.8) as claimed.

10. *Uniqueness.* We lastly show that the function $b$ is the unique solution to (3.8) in the class of continuous functions $t \mapsto b(t)$ on $[0, T]$ satisfying $b(t) \geq h(t)$ for all $t$ in $[0, T]$.

Take any continuous function $c$ on $[0, T]$ which solves (3.8) and satisfies $c(t) \geq h(t)$ for all $t \in [0, T]$. Motivated by (3.49) above, define the continuous function $U^c \colon [0, T] \times \mathbb{R}_+ \to \mathbb{R}$ by

(3.51)
$$U^c(t, x) = \mathsf{E}_{t,x}(G(T, X_T))$$
$$- \mathsf{E}_{t,x}\left(\int_0^{T-t} H(t+r, X_{t+r}) I(X_{t+r} > c(t+r))\, dr\right)$$

and observe that $c$ solving (3.8) means exactly that $U^c(t, c(t)) = G(t, c(t))$ for all $t \in [0, T]$. Let $D_c := \{(t, x) \in [0, T] \times \mathbb{R}_+ \mid x \geq c(t)\}$ so that $D_c$ is closed and plays the role of a "stopping set" for $c$. To avoid confusion we will denote by $D_b$ the original stopping set from (3.5) defined by the function $b$.

(i) We show that $U^c = G$ on $D_c$. Since $X$ is Markov, the process

(3.52) $$U^c(t+s, X_{t+s}) - \int_0^s H(t+r, X_{t+r}) I(X_{t+r} > c(t+r))\, dr$$

is a martingale under $\mathsf{P}_{t,x}$ for all $s \in [0, T-t]$. Take any point $(t, x) \in D_c$ and consider the stopping time

(3.53) $$\sigma_c = \inf\{s \in [0, T-t] \mid (t+s, X_{t+s}) \notin D_c\}$$

under the measure $\mathsf{P}_{t,x}$. Since $U^c(t, c(t)) = G(t, c(t))$ for all $t \in [0, T]$ and $U^c(T, x) = G(T, x)$ for all $x \in \mathbb{R}_+$, we must have $U^c(t + \sigma_c, X_{t+\sigma_c}) = G(t + \sigma_c, X_{t+\sigma_c})$. Inserting $\sigma_c$ in (3.52), taking $\mathsf{P}_{t,x}$ expectations and using the optional sampling theorem [recalling (3.29) together with (3.11) and (3.12) above] we find that

$$U^c(t, x) = \mathsf{E}_{t,x}(U^c(t + \sigma_c, X_{t+\sigma_c}))$$
$$- \mathsf{E}_{t,x}\left(\int_0^{\sigma_c} H(t+r, X_{t+r}) I((t+r, X_{t+r}) \in D_c)\, dr\right)$$

(3.54)
$$= \mathsf{E}_{t,x}(G(t+\sigma_c, X_{t+\sigma_c})) - \mathsf{E}_{t,x}\left(\int_0^{\sigma_c} H(t+r, X_{t+r})\,dr\right)$$
$$= G(t,x),$$

where in the last equality we used (3.14). This shows that $U^c = G$ on $D_c$ as claimed.

(ii) We show that $U^c(t,x) \geq V(t,x)$ for all $(t,x) \in [0,T] \times \mathbb{R}_+$. To see this take any $(t,x) \in [0,T] \times \mathbb{R}_+$ and consider the stopping time

(3.55) $$\tau_c = \inf\{s \in [0, T-t] \mid (t+s, X_{t+s}) \in D_c\}$$

under $\mathsf{P}_{t,x}$. We claim that $U^c(t+\tau_c, X_{t+\tau_c}) = G(t+\tau_c, X_{t+\tau_c})$. Indeed, if $(t,x) \in D_c$, then $\tau_c = 0$ so that $U^c(t,x) = G(t,x)$ by the argument above. Conversely if $(t,x) \notin D_c$, then the result follows since $U^c(t, c(t)) = G(t, c(t))$ for all $t \in [0, T]$ and $U^c(T, x) = G(T, x)$ for all $x \in \mathbb{R}_+$. Inserting $\tau_c$ in (3.52) and using the optional sampling theorem, we see that

(3.56)
$$U^c(t,x) = \mathsf{E}_{t,x}(U(t+\tau_c, X_{t+\tau_c}))$$
$$- \mathsf{E}_{t,x}\left(\int_0^{\tau_c} H(t+s, X_{t+s})I((t+s, X_{t+s}) \in D_c)\,ds\right)$$
$$= \mathsf{E}_{t,x}(G(t+\tau_c, X_{t+\tau_c})) \geq V(t,x),$$

where the second identity follows from the definition of $\tau_c$. We conclude that $U^c \geq V$ on $[0,T] \times \mathbb{R}_+$ as claimed.

(iii) We show that $D_b \subseteq D_c$. Suppose this is not the case so that there exists some time $t \in [0,T)$ at which $b(t) < c(t)$. Choose any $x > c(t)$ and consider the stopping time

(3.57) $$\sigma_b = \inf\{s \in [0, T-t] \mid (t+s, X_{t+s}) \notin D_b\}$$

under the measure $\mathsf{P}_{t,x}$. Replacing $s$ with $\sigma_b$ in (3.48) and (3.52) and using the optional sampling theorem we find

(3.58) $$\mathsf{E}_{t,x}(V(t+\sigma_b, X_{t+\sigma_b})) = V(t,x) + \mathsf{E}_{t,x}\left(\int_0^{\sigma_b} H(t+s, X_{t+s})\,ds\right),$$

(3.59)
$$\mathsf{E}_{t,x}(U^c(t+\sigma_b, X_{t+\sigma_b}))$$
$$= U^c(t,x) + \mathsf{E}_{t,x}\left(\int_0^{\sigma_b} H(t+s, X_{t+s})I((t+s, X_{t+s}) \in D_c)\,ds\right).$$

Since $(t,x)$ belongs to both $D_b$ and $D_c$ it follows that $U^c(t,x) = V(t,x) = G(t,x)$, and the fact that $U^c(t+\sigma_b, X_{t+\sigma_b}) \geq V(t+\sigma_b, X_{t+\sigma_b}) = G(t+\sigma_b, X_{t+\sigma_b})$ implies

(3.60) $$\mathsf{E}_{t,x}\left(\int_0^{\sigma_b} H(t+s, X_{t+s})I((t+s, X_{t+s}) \notin D_c)\,ds\right) \leq 0.$$

The assumption that $b(t) < c(t)$ together with the continuity of the functions $c$ and $b$ means that there exists a small enough $u \in (t, T]$ such that $b(s) < c(s)$ for all $s \in [t, u]$. Consequently the $\mathsf{P}_{t,x}$ probability of $X$ spending a strictly positive amount of time (w.r.t. Lebesgue measure) in this set is strictly positive. Combined with the fact that $b$ lies above $c$, this forces the expectation above to be strictly positive and provides a contradiction.

(iv) We show that $D_c = D_b$. Suppose that this is not the case so that $c(t) < b(t)$ for some $t \in [0, T]$. Choose any point $x \in (c(t), b(t))$ and consider the stopping time

$$\tau_D = \inf\{s \in [0, T-t] \mid (t+s, X_{t+s}) \in D_b\} \tag{3.61}$$

under $\mathsf{P}_{t,x}$. Inserting $\tau_D$ in (3.48) and (3.52), taking $\mathsf{P}_{t,x}$ expectations and using the optional sampling theorem we obtain

$$\mathsf{E}_{t,x}(G(t+\tau_D, X_{t+\tau_D})) = V(t,x), \tag{3.62}$$

$$\begin{aligned}\mathsf{E}_{t,x}(U^c(t+\tau_D, X_{t+\tau_D})) \\ = U^c(t,x) + \mathsf{E}_{t,x}\left(\int_0^{\tau_D} H(t+s, X_{t+s})I((t+s, X_{t+s}) \in D_c)\, ds\right).\end{aligned} \tag{3.63}$$

Since $D_b \subseteq D_c$ and $U^c$ equals $G$ on $D_c$ we must have $U^c(t+\tau_D, X_{t+\tau_D}) = G(t+\tau_D, X_{t+\tau_D})$, and using the fact that $U^c \geq V$ we find that

$$\mathsf{E}_{t,x}\left(\int_0^{\tau_D} H(t+s, X_{t+s})I((t+s, X_{t+s}) \in D_c)\, ds\right) \leq 0. \tag{3.64}$$

However, as before the continuity of the functions $b$ and $c$ combined with the fact that $c$ lies above $h$ forces the expectation to be strictly positive and provides a contradiction. We therefore conclude that $c(t) = b(t)$ for all $t \in [0, T]$ completing the proof. □

**4. The supremum problem.** We now consider the supremum formulation (1.4) of the stock selling problem. We recall from Section 2 that $B = (B_t)_{0 \leq t \leq T}$ denotes a standard Brownian motion, $B^\lambda = (B_t^\lambda)_{0 \leq t \leq T}$ denotes Brownian motion with drift $\lambda \in \mathbb{R}$ defined by $B_t^\lambda = B_t + \lambda t$ for $t \in [0, T]$, and $S^\lambda = (S_t^\lambda)_{0 \leq t \leq T}$ denotes its running maximum process, that is, $S_t^\lambda = \max_{0 \leq s \leq t} B_s^\lambda$ for $t \in [0, T]$. By (2.1)–(2.4) above we see that the problem (1.4) is equivalent to

$$V_2 = \sup_{0 \leq \tau \leq T} \mathsf{E}\left(\frac{Z_\tau}{M_T}\right) = \sup_{0 \leq \tau \leq T} \mathsf{E}(e^{\sigma(B_\tau^\lambda - S_T^\lambda)}), \tag{4.1}$$

where $\lambda = (\mu - \sigma^2/2)/\sigma$ with $\sigma > 0$ and $\mu \in \mathbb{R}$ given and fixed, and the supremum is taken over all stopping times $\tau$ of $Z$ (or $B$ equivalently) taking values in $[0, T]$.

1. As mentioned in the Introduction, it is surprising that this optimal prediction problem turns out to have a solution which is quite different from the solution to the infimum formulation when $\mu \in (0, \sigma^2)$. Indeed, it was shown in [15] that the function $H = G_t - \lambda G_x + \frac{1}{2} G_{xx}$ in the supremum formulation is strictly positive when $\mu \geq \sigma^2$ (i.e., $\lambda \geq \sigma/2$) and strictly negative when $\mu \leq 0$ (i.e., $\lambda \leq -\sigma/2$). This global argument implies that (4.2) holds when $\lambda \geq \sigma/2$ and that (4.3) holds when $\lambda \leq -\sigma/2$. However, when $\mu \in (0, \sigma^2)$ [i.e., $\lambda \in (-\sigma/2, \sigma/2)$] the function $H$ may take on both positive and negative values and the same global argument is no longer applicable (see Remark 3 for more details). Moreover, in view of the fact that the optimal stopping boundary in the infimum formulation of the problem is nontrivial in this case (recall Theorem 1 above) one could expect that the same fact holds in the supremum formulation as well. We now show, however, that this is not the case. Indeed, the probabilistic proof presented below applies to all cases of $\mu \in \mathbb{R}$ simultaneously, resolves the problem when $\mu \in (0, \sigma^2)$, and reveals the "bang–bang" character of the optimal strategy at $\mu = \sigma^2/2$. In the version of [15] that we received after communicating this proof, it was shown that when $\mu \in [\sigma^2/2, \sigma^2)$ it is more optimal to continue to the final time $T$ than to stop at once. When combined with the general result from optimal stopping theory (after verifying sufficient conditions) that the supremum is attained at the first entry time to the set where the value and gain functions are equal, this fact also yields the inequality (4.2) for all $\lambda \geq 0$. Finally, the inequalities (4.2) and (4.3) are interesting in their own right and rather unusual: they are, to the best of our knowledge, the first time such inequalities involving the maximum at a future time have appeared.

THEOREM 2. *Consider the optimal prediction problem (4.1). If $\lambda \geq 0$ then*

$$\mathsf{E}(e^{\sigma(B_\tau^\lambda - S_T^\lambda)}) \leq \mathsf{E}(e^{\sigma(B_T^\lambda - S_T^\lambda)}) \tag{4.2}$$

*for all stopping times $\tau$ of $B$ taking values in $[0, T]$. If $\lambda \leq 0$ then*

$$\mathsf{E}(e^{\sigma(B_\tau^\lambda - S_T^\lambda)}) \leq \mathsf{E}(e^{-\sigma S_T^\lambda}) \tag{4.3}$$

*for all stopping times $\tau$ of $B$ taking values in $[0, T]$. This shows that the optimal stopping time $\tau_*$ in (4.1) is described by the following "bang–bang" rule: when $\mu \geq \sigma^2/2$ then $\tau_* \equiv T$, and when $\mu \leq \sigma^2/2$ then $\tau_* \equiv 0$. (Note that when $\mu = \sigma^2/2$ then both $\tau_* \equiv T$ and $\tau_* \equiv 0$ are optimal, while in all other cases $\tau_*$ is $\mathsf{P}$-a.s. unique.)*

PROOF. Observe from (4.1) and the scaling property of Brownian motion that there is no restriction in assuming that $\sigma = 1$ if we likewise adjust the terminal time $T$ accordingly.

1. We first consider the case when $\lambda \geq 0$ (i.e., $\mu \geq \sigma^2/2$). In order to prove that $\tau_* \equiv T$ we need to show that (4.2) holds with $\sigma = 1$, that is,

$$\mathsf{E}(e^{B_T^\lambda - S_T^\lambda}) \geq \mathsf{E}(e^{B_\tau^\lambda - S_T^\lambda}) \tag{4.4}$$

for every stopping time $\tau$ of $B$ taking values in $[0, T]$. Clearly (4.4) will follow from

$$\text{(4.5)} \quad \mathsf{E}(e^{B_T^\lambda - S_T^\lambda} \mid \mathcal{F}_\tau^B) \geq \mathsf{E}(e^{B_\tau^\lambda - S_T^\lambda} \mid \mathcal{F}_\tau^B)$$

being valid for all stopping times $\tau$ of $B$ taking values in $[0, T]$, where $(\mathcal{F}_t^B)_{0 \leq t \leq T}$ is the natural filtration generated by $B$. To prove (4.5) it is enough to show that

$$\text{(4.6)} \quad \mathsf{E}(e^{B_T^\lambda - S_T^\lambda} \mid \mathcal{F}_t^B) \geq \mathsf{E}(e^{B_t^\lambda - S_T^\lambda} \mid \mathcal{F}_t^B)$$

for all $t \in [0, T]$, since if (4.6) were to hold pointwise between these two (continuous) processes for all times $t \in [0, T]$, it would hold for all stopping times as well. To establish (4.6), fix any $t \in [0, T]$ and note by the stationary independent increments of $B^\lambda$ that

$$\text{(4.7)} \quad \begin{aligned} &\mathsf{E}(\exp(B_T^\lambda - S_T^\lambda) \mid \mathcal{F}_t^B) \\ &= \mathsf{E}\left(\exp\left(B_T^\lambda - S_t^\lambda \vee \max_{t \leq s \leq T}(B_s^\lambda - B_t^\lambda + B_t^\lambda)\right) \mid \mathcal{F}_t^B\right) \\ &= \mathsf{E}\left(\exp\left(B_T^\lambda - B_t^\lambda - (S_t^\lambda - B_t^\lambda) \vee \max_{t \leq s \leq T}(B_s^\lambda - B_t^\lambda)\right) \mid \mathcal{F}_t^B\right) \\ &= \mathsf{E}(e^{B_{T-t}^\lambda - x \vee S_{T-t}^\lambda})|_{x = S_t^\lambda - B_t^\lambda}. \end{aligned}$$

Similarly we find that

$$\text{(4.8)} \quad \mathsf{E}(e^{B_t^\lambda - S_T^\lambda} \mid \mathcal{F}_t^B) = \mathsf{E}(e^{-x \vee S_{T-t}^\lambda})|_{x = S_t^\lambda - B_t^\lambda}$$

and so from (4.7) and (4.8) we see that for (4.6) it is enough to show that

$$\text{(4.9)} \quad \mathsf{E}(e^{B_t^\lambda - x \vee S_t^\lambda}) \geq \mathsf{E}(e^{-x \vee S_t^\lambda})$$

for all $x \geq 0$ and $t \in [0, T]$ whenever $\lambda \geq 0$. To derive (4.9) we make the key observation that (4.9) holds for all $\lambda \geq 0$ if and only if it holds for $\lambda = 0$. Indeed suppose that (4.9) holds for $\lambda = 0$ and let $\lambda > 0$ be given and fixed. Then by the assumption we see that

$$\text{(4.10)} \quad \begin{aligned} \mathsf{E}(e^{B_t^\lambda - x \vee S_t^\lambda}) &\geq \mathsf{E}(e^{B_t + \lambda t - x \vee (S_t + \lambda t)}) \geq \mathsf{E}(e^{B_t + \lambda t - (x \vee S_t + \lambda t)}) \\ &= \mathsf{E}(e^{B_t - x \vee S_t}) \geq \mathsf{E}(e^{-x \vee S_t}) \geq \mathsf{E}(e^{-x \vee S_t^\lambda}) \end{aligned}$$

proving the claim. Thus it is enough to show that

$$\text{(4.11)} \quad \mathsf{E}(e^{B_t - x \vee S_t}) \geq \mathsf{E}(e^{-x \vee S_t})$$

for all $x \geq 0$ and $t \in [0, T]$. To derive (4.11) recall that $x \vee S_t - B_t =^{\text{law}} |x + B_t|$ and $S_t =^{\text{law}} |B_t|$. It follows then that (4.11) can be written as

$$\text{(4.12)} \quad \mathsf{E}(e^{-|x + B_t|}) \geq \mathsf{E}(e^{-(x \vee |B_t|)})$$

which by the scaling property of Brownian motion is the same as

$$\mathsf{E}(e^{-\sqrt{t}|(x/\sqrt{t})+B_1|}) \geq \mathsf{E}(e^{-\sqrt{t}((x/\sqrt{t})\vee|B_1|)}). \tag{4.13}$$

Therefore it is enough to show that

$$\mathsf{E}(e^{-c|x+B_1|}) \geq \mathsf{E}(e^{-c(x\vee|B_1|)}) \tag{4.14}$$

for all $c \geq 0$ and all $x \geq 0$. Setting $\varphi(x) = (1/\sqrt{2\pi})e^{-x^2/2}$ it is easy to see that the left-hand side in (4.14) equals

$$L = e^{cx} \int_{-\infty}^{-x} e^{cy}\varphi(y)\,dy + e^{-cx} \int_{-x}^{x} e^{-cy}\varphi(y)\,dy \\ + e^{-cx} \int_{x}^{\infty} e^{-cy}\varphi(y)\,dy \tag{4.15}$$

while the right-hand side is given by

$$R = \int_{-\infty}^{-x} e^{cy}\varphi(y)\,dy + e^{-cx}\int_{-x}^{x}\varphi(y)\,dy + \int_{x}^{\infty} e^{-cy}\varphi(y)\,dy. \tag{4.16}$$

Turning to (4.15), setting $y = -z$ in the first integral and adding this to the last integral, and doing likewise in (4.16), we see that the resulting expressions satisfy

$$(e^{cx} + e^{-cx}) \int_{x}^{\infty} e^{-cy}\varphi(y)\,dy \geq 2 \int_{x}^{\infty} e^{-cy}\varphi(y)\,dy \tag{4.17}$$

since $\frac{1}{2}(e^{cx}+e^{-cx}) = \operatorname{ch}(cx) \geq 1$ for all $c \geq 0$ and all $x \geq 0$. Therefore to complete the proof it is enough to show that

$$f(x) := \int_{-x}^{x} e^{-cy}\varphi(y)\,dy \geq \int_{-x}^{x} \varphi(y)\,dy =: g(x) \tag{4.18}$$

for all $c \geq 0$ and $x \geq 0$. For this, note that $f(0) = g(0) = 0$ and

$$f'(x) = (e^{-cx} + e^{cx})\varphi(x) \geq 2\varphi(x) = g'(x) \tag{4.19}$$

for all $x \geq 0$ since $\operatorname{ch}(cx) \geq 1$ for all $c \geq 0$ and $x \geq 0$. Thus $f(x) \geq g(x)$ for all $x \geq 0$ and the proof of (4.2) is complete. Note also that when $\lambda > 0$ the proof above shows that equality in (4.2) is not attained at any other stopping time $\tau$ of $B$ with values in $[0,T]$ such that $\mathsf{P}(\tau < T) > 0$. Thus $\tau_* \equiv T$ is the only optimal stopping time (P-a.s.) when $\mu > \sigma^2/2$.

2. We next consider the case when $\lambda \leq 0$ (i.e., $\mu \leq \sigma^2/2$). In order to prove that $\tau_* \equiv 0$ we need to show that (4.3) holds with $\sigma = 1$ (by Brownian scaling). Writing the drift as $-\lambda$ for $\lambda \geq 0$, we see that the problem reduces to showing that

$$\mathsf{E}(e^{-S_T^{-\lambda}}) \geq \mathsf{E}(e^{B_\tau^{-\lambda} - S_T^{-\lambda}}) \tag{4.20}$$

for every stopping time $\tau$ of $B$ taking values in $[0,T]$. This inequality is more involved than the inequality (4.4) above (when $\lambda \in \{0\} \cup [1/2, \infty)$ a shorter proof can be given using Girsanov's theorem).

We begin by rewriting the left-hand side of (4.20). For this, recall that $S_T^{-\lambda} =^{\text{law}} B_T^{-\lambda} - I_T^{-\lambda} =^{\text{law}} -B_T^\lambda + S_T^\lambda$, where $I_T^{-\lambda} = \inf_{0 \leq t \leq T} B_t^{-\lambda}$, so that (4.20) reads

$$(4.21) \qquad \mathsf{E}(e^{B_T^\lambda - S_T^\lambda}) \geq \mathsf{E}(e^{B_\tau^{-\lambda} - S_T^{-\lambda}}),$$

where $\tau$ and $\lambda$ are as above. Clearly it is enough to show that

$$(4.22) \qquad \mathsf{E}(e^{B_T^\lambda - S_T^\lambda} \mid \mathcal{F}_\tau^B) \geq \mathsf{E}(e^{B_\tau^{-\lambda} - S_T^{-\lambda}} \mid \mathcal{F}_\tau^B)$$

for all $\tau$ as above, and as at (4.6) above, the inequality (4.22) will follow if

$$(4.23) \qquad \mathsf{E}(e^{B_T^\lambda - S_T^\lambda} \mid \mathcal{F}_t^B) \geq \mathsf{E}(e^{B_t^{-\lambda} - S_T^{-\lambda}} \mid \mathcal{F}_t^B)$$

for all $t \in [0,T]$. For this, by the same arguments as in (4.7) above we find that

$$(4.24) \qquad \mathsf{E}(e^{B_T^\lambda - S_T^\lambda} \mid \mathcal{F}_t^B) = \mathsf{E}(e^{B_{T-t}^\lambda - x \vee S_{T-t}^\lambda})\big|_{x = S_t^\lambda - B_t^\lambda}$$

and similarly

$$(4.25) \qquad \mathsf{E}(e^{B_t^{-\lambda} - S_T^{-\lambda}} \mid \mathcal{F}_t^B) = \mathsf{E}(e^{-x \vee S_{T-t}^{-\lambda}})\big|_{x = S_t^{-\lambda} - B_t^{-\lambda}}$$

for all $t \in [0,T]$. Since $\lambda \geq 0$ we have that

$$(4.26) \qquad \begin{aligned} S_t^\lambda - B_t^\lambda &\leq S_t + \lambda t - (B_t + \lambda t) \\ &= S_t - B_t = S_t - \lambda t - (B_t - \lambda t) \leq S_t^{-\lambda} - B_t^{-\lambda} \end{aligned}$$

for all $t \in [0,T]$. Using (4.26) in (4.24) we see that

$$(4.27) \qquad \mathsf{E}(e^{B_T^\lambda - S_T^\lambda} \mid \mathcal{F}_t^B) \geq \mathsf{E}(e^{B_{T-t}^\lambda - x \vee S_{T-t}^\lambda})\big|_{x = S_t^{-\lambda} - B_t^{-\lambda}}$$

so that (4.23) will follow if we are able to show that

$$(4.28) \qquad \mathsf{E}(e^{B_t^\lambda - x \vee S_t^\lambda}) \geq \mathsf{E}(e^{-x \vee S_t^{-\lambda}})$$

for all $x \geq 0$ and all $t \in [0,T]$. Note that the inequality (4.28) is sharper than the inequality (4.9) above since the right-hand side in (4.28) is larger, and once (4.28) is proved we will get (4.9) as a consequence. Note also that a similar approach based on (4.10) above is no longer possible in the case of (4.28). Lastly, observe that (4.28) is satisfied if either $\lambda = 0$ or $x = 0$. Indeed, the former case reduces to (4.9) above, while the latter case follows from the fact that $B_t^\lambda - S_t^\lambda =^{\text{law}} I_t^\lambda =^{\text{law}} -S_t^{-\lambda}$ so that (4.28) holds in particular.

To derive (4.28) in general, let us first note that $x \vee S_t^\lambda = x + (S_t^\lambda - x)^+$, where $\lambda$ can be positive or negative, so that (4.28) reads

$$\mathsf{E}(e^{B_t^\lambda - (S_t^\lambda - x)^+}) \geq \mathsf{E}(e^{-(S_t^{-\lambda} - x)^+}). \tag{4.29}$$

Recalling that (4.29) holds for $x = 0$, we see that it is enough to show that

$$\frac{\partial}{\partial x} \mathsf{E}(e^{B_t^\lambda - (S_t^\lambda - x)^+}) \geq \frac{\partial}{\partial x} \mathsf{E}(e^{-(S_t^{-\lambda} - x)^+}) \tag{4.30}$$

for all $x \geq 0$ and all $t \in [0, T]$. Interchanging $\frac{\partial}{\partial x}$ and $\mathsf{E}$ in (4.30) (which is easily justified by standard means) we see that (4.30) becomes

$$\mathsf{E}(e^{B_t^\lambda - (S_t^\lambda - x)} I(S_t^\lambda > x)) \geq \mathsf{E}(e^{-(S_t^{-\lambda} - x)} I(S_t^{-\lambda} > x)), \tag{4.31}$$

which upon multiplying by $e^{-x}$ is the same as

$$\mathsf{E}(e^{B_t^\lambda - S_t^\lambda} I(S_t^\lambda > x)) \geq \mathsf{E}(e^{-S_t^{-\lambda}} I(S_t^{-\lambda} > x)). \tag{4.32}$$

Since $(S_t^\lambda, S_t^\lambda - B_t^\lambda) =^{\text{law}} (B_t^\lambda - I_t^\lambda, -I_t^\lambda) =^{\text{law}} (-B_t^{-\lambda} + S_t^{-\lambda}, S_t^{-\lambda})$ we see that the left-hand side of (4.32), and thus (4.32) itself, can be rewritten as

$$\mathsf{E}(e^{-S_t^{-\lambda}} I(S_t^{-\lambda} - B_t^{-\lambda} > x)) \geq \mathsf{E}(e^{-S_t^{-\lambda}} I(S_t^{-\lambda} > x)). \tag{4.33}$$

Recall [see (3.2) above] that the density function $f$ of $(B_t^{-\lambda}, S_t^{-\lambda})$ is given explicitly by

$$f(t, b, s) = \sqrt{\frac{2}{\pi}} \frac{(2s - b)}{t^{3/2}} e^{-(2s-b)^2/(2t) - \lambda(b + \lambda t/2)} \tag{4.34}$$

for $s \geq 0$ and $b \leq s$. Using $f$ we can rewrite (4.33) as follows:

$$\int_0^\infty e^{-s} \int_{-\infty}^{s-x} f(t, b, s) \, db \, ds \geq \int_x^\infty e^{-s} \int_{-\infty}^s f(t, b, s) \, db \, ds. \tag{4.35}$$

Substituting $b' = b - (s - x)$ in the left-hand side, $s' = s - x$ and $b' = b - (s' + x)$ in the right-hand side, and setting

$$g(b, s) = (2s - b) e^{-(2s-b)^2/(2t) - \lambda b} \tag{4.36}$$

with $t \in [0, T]$ fixed, we find that (4.35) is equivalent to

$$\int_0^\infty e^{-s} \int_{-\infty}^0 g(b + s - x, s) \, db \, ds$$
$$\geq \int_0^\infty e^{-(s+x)} \int_{-\infty}^0 g(b + s + x, s + x) \, db \, ds. \tag{4.37}$$

Hence it is enough to show that

$$g(b + s - x, s) \geq g(b + s + x, s + x). \tag{4.38}$$

For this, note that we have

$$(4.39) \qquad g(b+s-x,s) = (s-b+x)e^{-(s-b+x)^2/(2t)-\lambda(b+s)+\lambda x},$$

$$(4.40) \qquad g(b+s+x,s+x) = (s-b+x)e^{-(s-b+x)^2/(2t)-\lambda(b+s)-\lambda x}.$$

A direct comparison of (4.39) and (4.40) shows that (4.38) holds for all $x \geq 0$ and this completes the proof. $\square$

REMARK 1 (*The "bang–bang" strategy*). To grasp the meaning of the "bang–bang" character of the optimal strategies in the supremum formulation (4.1), let us consider the optimal stopping problem

$$(4.41) \qquad V = \sup_{0 \leq \tau \leq T} \mathsf{E}(Z_\tau),$$

where the supremum is taken over all stopping times $\tau$ of the geometric Brownian motion $Z$ with drift $\mu \in \mathbb{R}$ and volatility $\sigma > 0$. By the sub/super/martingale property of $Z$ we see that the optimal stopping time $\tau_*$ is described by the following "bang–bang" rule: when $\mu \geq 0$ then $\tau_* \equiv T$ and when $\mu \leq 0$ then $\tau_* \equiv 0$. This shows that if we are to maximize the mean of $Z_\tau$ with reference to 1 [in the sense that $Z_\tau$ in (4.41) equals $Z_\tau/1$], then the "critical" drift $\mu_*$ equals 0. On the other hand, if this reference point is being replaced by the more ambitious reference point of the ultimate maximum $M_T$, then the result of Theorem 2 shows that the drifts $\mu$ belonging to $(0, \sigma^2/2)$ are no longer good enough for continuation and the "critical" drift equals $\sigma^2/2$ in this case.

Quite similarly, to relate these interpretations to the optimal strategies in the infimum formulation (2.4), let us consider the optimal stopping problem

$$(4.42) \qquad V = \inf_{0 \leq \tau \leq T} \mathsf{E}\left(\frac{1}{Z_\tau}\right),$$

where the infimum is taken over all stopping times $\tau$ of $Z$ as in (4.41) above. By the super/sub/martingale property of $Z$ we see that the optimal stopping time $\tau_*$ is described by the following "bang–bang" rule: when $\mu \geq \sigma^2$ then $\tau_* \equiv T$ and when $\mu \leq \sigma^2$ then $\tau_* \equiv 0$. This shows that if we are to penalize the mean of $1/Z_\tau$ for small $Z_\tau$ with reference to 1 [in the sense that the integrand in (4.42) equals $1/Z_\tau$], then the "critical" drift $\mu_*$ equals $\sigma^2$. On the other hand, if this reference point is replaced by the more ambitious reference point of the ultimate maximum $M_T$, then the result of Theorem 1 shows that the drifts $\mu$ belonging to $(0, \sigma^2)$ are no longer good enough for stopping, and the "critical" drift $\mu_*$ is diffused into a nontrivial function of time specified in Theorem 1.

REMARK 2 (*Implementing the "bang–bang" strategy*). As we have shown in Theorem 2 above, the optimal strategy in the supremum formulation of the stock selling problem is of the so-called "bang–bang" type: when $\mu \leq \sigma^2/2$ one stops immediately, and when $\mu > \sigma^2/2$ one waits until the final time $T$. There are a number of interesting questions which emerge when one considers how this strategy could be implemented in practice. One way is to exploit the "bang–bang" structure and leverage the fact that a single point (namely $\sigma^2/2$) defines the boundary between one simple strategy (sell the stock) and another (hold the stock).

Imagine an investor who owns a stock $Z$ following a geometric Brownian motion with unknown drift $\mu$ and known volatility $\sigma$. The investor has a history of observations and wishes to use this information to determine at which time before $T$ to sell the stock so as to maximize $\mathsf{E}(Z_\tau/M_T)$. Unfortunately this involves estimating the drift, and it is well known that this is particularly difficult and requires a prodigious amount of data to achieve with any kind of accuracy. Moreover, the optimal strategy is very sensitive to errors in the estimated value $\hat{\mu}$ of $\mu$ when close to $\sigma^2$. One approach which lends itself to engineering applications is to link this problem with two well-known problems from mathematical statistics:

(i) *Sequential testing.* To use the sequential testing approach, one assumes that the stock-price drift takes one of two possible values: $\mu_0 > \sigma^2/2$ or $\mu_1 \leq \sigma^2/2$. The aim is to test the null hypothesis $H_0 : \mu = \mu_0$ against the alternate hypothesis $H_1 : \mu = \mu_1$, and if $H_0$ is rejected one sells the stock. The test is performed by monitoring the process $f(Z_t)$ for a specified functional $f$ when $t$ runs from 0 to $T$, and stopping at the first time $\tau_*$ at which $f(Z_{\tau_*})$ belongs to a specified set $D_0 \cup D_1$. If $f(Z_{\tau_*})$ belongs to $D_0$, then one rejects $H_0$ and sells the stock, and if $f(Z_{\tau_*})$ belongs to $D_1$, then one does not reject $H_0$ and holds the stock until time $T$. For further information about the test and other ramifications in this direction see, for example, [12], Section 21.

(ii) *Quickest detection.* To use the quickest detection approach, one assumes that the stock-price drift is equal to $\mu_0 > \sigma^2/2$, and that at some independent (e.g., exponentially distributed) time $\theta$, the drift will change to $\mu_1 \leq \sigma^2/2$. The aim is to detect $\theta$ as quickly and as accurately as possible, and at this point to sell the stock. The test is performed by monitoring the process $f(Z_t)$ for a specified functional $f$ when $t$ runs from 0 to $T$, and stopping at the first time $\tau_*$ at which $f(Z_{\tau_*})$ belongs to a specified set $D$. At this point one sells the stock. For further information about the test and other ramifications in this direction see, for example, [12], Section 22.

REMARK 3. We chose to solve the optimal prediction problem (4.1) directly, proving that $\tau_* \equiv T$ or $\tau_* \equiv 0$ by establishing the inequalities (4.2) and (4.3). However, one can also tackle (4.1) with the same machinery as was used to solve the infimum formulation (1.3), and doing so reveals why the supremum formulation is inherently more complex than the infimum formulation. Calculations similar to those at Lemma 1 show that

$$V(t,x) := \sup_{0 \leq \tau \leq T-t} \mathsf{E}_{t,x}\left(\frac{Z_\tau}{M_T}\right) = \sup_{0 \leq \tau \leq T-t} \mathsf{E}_{t,x}(G(t+\tau, S^\lambda_{t+\tau} - B^\lambda_{t+\tau}))$$
$$= \sup_{0 \leq \tau \leq T-t} \mathsf{E}(G(t+\tau, X^x_\tau)) \quad (4.43)$$

for all $(t,x) \in [0,T] \times \mathbb{R}_+$, where the function $G$ above is equal to $G$ from (2.7) and (2.8) with $\sigma$ replaced by $-\sigma$ and $X^x = (X^x_t)_{0 \leq t \leq T} = (x \vee S^\lambda_t - B^\lambda_t)_{0 \leq t \leq T}$ for any $x \geq 0$.

Applying Itô's formula to $G$ [as at (3.14) above] and taking expectations, one finds by the optional sampling theorem [as at (3.16) above] that

$$(4.44) \qquad V(t,x) = G(t,x) + \sup_{0 \leq \tau \leq T-t} \mathsf{E}\left(\int_0^\tau H(t+s, X^x_s)\, ds\right)$$

for all $(t,x) \in [0,T] \times \mathbb{R}_+$, where $H = G_t - \lambda G_x + \frac{1}{2} G_{xx}$ is equal to $H$ from (3.1) and (3.15) with $\sigma$ replaced by $-\sigma$. A direct examination of the function $H$ shows that when $\mu \leq 0$ we have $H(t,x) < 0$, and when $\mu \geq \sigma^2$ we have $H(t,x) > 0$ for all $(t,x) \in [0,T] \times \mathbb{R}_+$ (this was derived in [15]). Considerations similar to those at (3.17)–(3.19) then show that if $\mu \leq 0$, it is optimal to stop immediately, whereas if $\mu \geq \sigma^2$, it is optimal to wait until time $T$.

However, when $\mu \in (0, \sigma^2)$ there exists a continuous decreasing function $h$ on $[0,T]$ with $h(T) = 0$ such that $\{H < 0\} = \{(t,x) \in [0,T] \times \mathbb{R}_+ \mid x < h(t)\}$. The same arguments as at (3.19) then show that the area above $h$ will be part of the continuation set, while the area below $h$ *may* contain a stopping set. In addition, it is also not true that $H_t \leq 0$. This peculiar behavior of $H$ and $H_t$ leads to several complications; for example, it is much harder to show that the optimal stopping boundary is regular for the diffusion [compare with (3.35) and (3.36) above] without resorting to purely analytic methods coming from the theory of free-boundary problems. Moreover, while solving the optimal stopping problem (4.43) is quite demanding, it is not required in order to solve the optimal stopping problem (1.4) where the process $X$ starts at 0 at time 0. All that is needed is to show that $(0,0)$ is a stopping point when $\mu \leq \sigma^2/2$, and that no point in $[0,T) \times \mathbb{R}_+$ is in the stopping set when $\mu > \sigma^2/2$. For this reason we chose to present the current proof in Section 4 rather than to solve the problem (4.43) directly. It may also be noted that the inequality (4.11) is strict when $x > 0$ and $t > 0$ [since $\mathrm{ch}(cx) > 1$ for $cx \neq 0$] so that no point $(t,x) \in [0,T) \times (0,\infty)$ belongs to the stopping set when $\mu = \sigma^2/2$, that is, the optimal stopping boundary is given by $b(t) = 0$ for $t \in [0,T]$ in this case.

2. To conclude the paper we briefly return to the infimum formulation (2.4). Recall (see Theorem 1) that for this problem it is optimal to wait until the final time $T$ when $\mu \geq \sigma^2$ and it is optimal to stop immediately if $\mu \leq 0$. These results were derived in Section 3 by reducing (2.4) to an adapted optimal stopping problem, and then using stochastic calculus techniques to examine the two cases. We now give a direct proof of these facts based on the methods developed in the proof of Theorem 2 and the Girsanov theorem. Quite similarly, the Girsanov theorem can also be used to give simpler proofs of the inequalities (4.2) and (4.3) when either $\mu \geq \sigma^2$ or $\mu \leq 0$ respectively.

THEOREM 3. *Consider the optimal prediction problem (2.4). If $\lambda \geq \sigma/2$ then*

$$(4.45) \qquad \mathsf{E}(e^{\sigma(S_T^\lambda - B_\tau^\lambda)}) \geq \mathsf{E}(e^{\sigma(S_T^\lambda - B_T^\lambda)})$$

*for all stopping times $\tau$ of $B$ taking values in $[0,T]$. If $\lambda \leq -\sigma/2$ then*

$$(4.46) \qquad \mathsf{E}(e^{\sigma(S_T^\lambda - B_\tau^\lambda)}) \geq \mathsf{E}(e^{\sigma S_T^\lambda})$$

*for all stopping times $\tau$ of $B$ taking values in $[0,T]$. This shows that the optimal stopping time $\tau_*$ in (2.4) is described as follows: when $\mu \geq \sigma^2$ we have $\tau_* \equiv T$, and when $\mu \leq 0$ we have $\tau_* \equiv 0$. [Recalling that $\tau_*$ is nontrivial when $\mu \in (0, \sigma^2)$ we see that (4.45) and (4.46) fail to hold for all stopping times when either $\lambda < \sigma/2$ or $\lambda > -\sigma/2$ respectively.]*

PROOF. We will be rather brief as many of the ideas are very similar to the proof of Theorem 2. Recall that we can set $\sigma = 1$ by Brownian scaling.

1. We first consider (4.45) and observe, as at (4.6) above, that it is enough to show that

$$(4.47) \qquad \mathsf{E}(e^{S_T^\lambda - B_t^\lambda} \mid \mathcal{F}_t^B) \geq \mathsf{E}(e^{S_T^\lambda - B_T^\lambda} \mid \mathcal{F}_t^B)$$

for all $t \in [0,T]$. Recalling (4.7)–(4.9) above, we similarly see that (4.47) reduces to showing

$$(4.48) \qquad \mathsf{E}(e^{x \vee S_t^\lambda}) \geq \mathsf{E}(e^{x \vee S_t^\lambda - B_t^\lambda})$$

for all $t \in [0,T]$ and all $x \geq 0$. Although a similar trick as at (4.10) [reducing (4.48) for "large" $\lambda$ to (4.48) with "smaller" $\lambda$] still applies, one notes, however, that (4.48) fails for $\lambda = 0$. [This can be seen by mimicking the arguments from (4.11) onward.] On the other hand, the inequality (4.48) does hold for $\lambda = 1/2$. Indeed, by the Girsanov theorem we have

$$(4.49) \qquad \begin{aligned} \mathsf{E}(e^{x \vee S_t^{1/2} - B_t^{1/2}}) &= \mathsf{E}(e^{x \vee S_t^{1/2} - B_t - t/2}) = \widetilde{\mathsf{E}}(e^{x \vee S_t^{1/2}}) \\ &= \widetilde{\mathsf{E}}(e^{x \vee \widetilde{S}_t^{-1/2}}) \leq \mathsf{E}(e^{x \vee S_t^{1/2}}) \end{aligned}$$

since $B_t + \frac{t}{2} = B_t + t - t + \frac{t}{2} = \widetilde{B}_t - \frac{t}{2}$ for $t \in [0,T]$, where $\widetilde{B} = (\widetilde{B}_t)_{0 \le t \le T}$ is a standard Brownian motion under $\widetilde{P}$. Hence if $\lambda \ge 1/2$ then

$$
\begin{aligned}
\mathsf{E}(e^{x \vee S_t^\lambda - B_t^\lambda}) &= \mathsf{E}(e^{x \vee \max_{0 \le s \le t}(B_s + \lambda s) - B_t^\lambda}) \\
&= \mathsf{E}(e^{x \vee \max_{0 \le s \le t}(B_s + s/2 + (\lambda - 1/2)s) - B_t - \lambda t}) \\
&\le \mathsf{E}(e^{x \vee S_t^{1/2} + (\lambda - 1/2)t - B_t - \lambda t}) = \mathsf{E}(e^{x \vee S_t^{1/2} - B_t^{1/2}}) \\
&\le \mathsf{E}(e^{x \vee S_t^{1/2}}) \le \mathsf{E}(e^{x \vee S_t^\lambda})
\end{aligned}
\tag{4.50}
$$

for all $t \in [0,T]$. This establishes (4.48) and completes the proof of (4.45).

2. To prove (4.46) for all $\lambda \le -1/2$, we first write the drift as $-\lambda$ for $\lambda \ge 1/2$. Mimicking the arguments from (4.20)–(4.28) and writing $\mathsf{E}(e^{S_T^{-\lambda}}) = \mathsf{E}(e^{S_T^\lambda - B_T^\lambda})$, we see that it is enough to establish

$$
\mathsf{E}(e^{x \vee S_t^{-\lambda}}) \ge \mathsf{E}(e^{x \vee S_t^\lambda - B_t^\lambda}) \tag{4.51}
$$

for all $t \ge 0$ and all $x \ge 0$ whenever $\lambda \ge 1/2$. Performing the same computations as at (4.29)–(4.33) leads to the inequality

$$
\mathsf{E}(e^{S_t^{-\lambda}} I(S_t^{-\lambda} > x)) \le \mathsf{E}(e^{S_t^{-\lambda}} I(S_t^{-\lambda} - B_t^{-\lambda} > x)), \tag{4.52}
$$

which has to be proved. Recalling the function $f$ from (4.34), expanding the expectations as integrals and making the same substitutions as below (4.35), we can rewrite (4.52) as

$$
\begin{aligned}
\int_0^\infty e^s \int_{-\infty}^0 e^x g(b+s+x, s+x) \, db \, ds \\
\le \int_0^\infty e^s \int_{-\infty}^0 g(b+s-x, s) \, db \, ds,
\end{aligned}
\tag{4.53}
$$

where the function $g$ is given by (4.36) for $t \ge 0$ given and fixed. Since

$$
e^x g(b+s+x, s+x) = (s-b+x)e^{-(s-b+x)^2/(2t) - \lambda(b+s) + x(1-\lambda)}, \tag{4.54}
$$

$$
g(b+s-x, s) = (s-b+x)e^{-(s-b+x)^2/(2t) - \lambda(b+s) + \lambda x}, \tag{4.55}
$$

we see that (4.53) will be satisfied whenever $\lambda \ge 1/2$. This completes the proof. □

REMARK 4 (*Key inequalities*). A closer examination of the proofs in this section will reveal that there are two key inequalities that were derived in establishing Theorems 2 and 3. For the sake of completeness we list them here:

$$
\mathsf{E}(e^{B_t^\lambda - x \vee S_t^\lambda}) \ge \mathsf{E}(e^{-x \vee S_t^{-\lambda}}) \qquad \text{for all } \lambda \ge -1/2, \tag{4.56}
$$

$$
\mathsf{E}(e^{x \vee S_t^{-\lambda}}) \ge \mathsf{E}(e^{x \vee S_t^\lambda - B_t^\lambda}) \qquad \text{for all } \lambda \ge 1/2, \tag{4.57}
$$

whenever $t \geq 0$ and $x \geq 0$. In fact the inequality (4.56) was derived only for $\lambda \geq 0$ above; however, a careful examination of the calculations in (4.51)–(4.54) and (4.28)–(4.40) shows that it also holds for $\lambda \geq -1/2$. Moreover, one can likewise verify that

$$(4.58) \qquad \mathsf{E}(e^{B_t^\lambda - x \vee S_t^\lambda}) \leq \mathsf{E}(e^{-x \vee S_t^{-\lambda}}) \qquad \text{for all } \lambda \leq -1/2,$$

$$(4.59) \qquad \mathsf{E}(e^{x \vee S_t^{-\lambda}}) \leq \mathsf{E}(e^{x \vee S_t^\lambda - B_t^\lambda}) \qquad \text{for all } \lambda \leq 1/2,$$

whenever $t \geq 0$ and $x \geq 0$.

School of Mathematics
The University of Manchester
Oxford Road
Manchester M13 9PL
United Kingdom
E-mail: Jacques.Du-Toit@postgrad.manchester.ac.uk
Goran.Peskir@manchester.ac.uk